\documentclass{aa}

\graphicspath{{./}{plots/}}

\usepackage[normalem]{ulem}
\usepackage{soul}
\usepackage{natbib,twoopt, xcolor, amsmath, booktabs}
\usepackage[breaklinks=true]{hyperref} 
\bibpunct{(}{)}{;}{a}{}{,}             
\makeatletter
  \newcommandtwoopt{\citeads}[3][][]{\href{http://adsabs.harvard.edu/abs/#3}%
    {\def\hyper@linkstart##1##2{}%
     \let\hyper@linkend\@empty\citealp[#1][#2]{#3}}}
  \newcommandtwoopt{\citepads}[3][][]{\href{http://adsabs.harvard.edu/abs/#3}%
    {\def\hyper@linkstart##1##2{}%
     \let\hyper@linkend\@empty\citep[#1][#2]{#3}}}
  \newcommandtwoopt{\citetads}[3][][]{\href{http://adsabs.harvard.edu/abs/#3}%
    {\def\hyper@linkstart##1##2{}%
     \let\hyper@linkend\@empty\citet[#1][#2]{#3}}}
  \newcommandtwoopt{\citeyearads}[3][][]%
    {\href{http://adsabs.harvard.edu/abs/#3}
    {\def\hyper@linkstart##1##2{}%
     \let\hyper@linkend\@empty\citeyear[#1][#2]{#3}}}
\makeatother

\begin{document}



\title{Evolution of the radial ISM metallicity gradient in the Milky Way disk since redshift $\approx 3$}

\author{B. Ratcliffe\inst{1} 
\and S. Khoperskov\inst{1}
\and I. Minchev\inst{1}
\and N. D. Lee\inst{2,3}
\and T. Buck\inst{4,5}
\and L. Marques\inst{1}
\and L. Lu\inst{6,7}
\and M. Steinmetz\inst{1}}
    
\institute{Leibniz-Institut für Astrophysik Potsdam (AIP), An der Sternwarte 16, 14482 Potsdam, Germany
\and BIFOLD, Berlin Institute for the Foundations of Learning and Data, Berlin, Germany
\and Technische Universität Berlin, Berlin, Germany
\and Universit\"at Heidelberg, Interdisziplin\"ares Zentrum f\"ur Wissenschaftliches Rechnen, Im Neuenheimer Feld 205, D-69120 Heidelberg, Germany
\and Universit\"at Heidelberg, Zentrum f\"ur Astronomie, Institut f\"ur Theoretische Astrophysik, Albert-Ueberle-Straße 2, D-69120 Heidelberg, Germany
\and Department of Astronomy, The Ohio State University, Columbus, 140 W 18th Ave, OH 43210, USA
\and Center for Cosmology and Astroparticle Physics (CCAPP), The Ohio State University, 191 W. Woodruff Ave., Columbus, OH 43210, USA}

\date{Received \today; accepted ...}

\abstract
{Recent works identified a way to recover the time evolution of a galaxy's disk metallicity gradient from the shape of its age--metallicity relation. However, the success of the method is dependent on how the width of the star-forming region evolves over time, which in turn is dependent on a galaxy's present-day bar strength.}
{In this paper, we account for the time variation in the width of the star-forming region when deriving the interstellar medium (ISM) metallicity gradient evolution over time ($\rm \nabla [Fe/H](\tau)$), which provides more realistic birth radii estimates of Milky Way (MW) disk stars.}
{Using MW/Andromeda analogues from the TNG50 simulation, we quantified the disk growth of newly born stars as a function of present-day bar strength to provide a correction that improves recovery of $\rm \nabla [Fe/H](\tau)$.}
{In TNG50, we find that our correction reduces the median absolute error in recovering $\rm \nabla [Fe/H] (\tau)$ by over 30\%. To confirm its universality, we test our correction on two galaxies from NIHAO-UHD and find the median absolute error is over 3 times smaller even in the presence of observational uncertainties for the barred, MW-like galaxy. Applying our correction to APOGEE DR17 red giant MW disk stars suggests the effects of merger events on $\rm \nabla [Fe/H](\tau)$ are less significant than originally found, and the corresponding estimated birth radii expose epochs when different migration mechanisms dominated.}
{Our correction to account for the growth of the star-forming region in the disk allows for better recovery of the evolution of the MW disk's ISM metallicity gradient and, thus, more meaningful stellar birth radii estimates. With our results, we are able to suggest the evolution of not only the ISM gradient, but also the total stellar disk radial metallicity gradient, providing key constraints to select MW analogues across redshift.}

\keywords{Stars: abundances, Galaxy: disk, Galaxies: evolution}
\authorrunning{Ratcliffe et al.}
\titlerunning{}
\maketitle

\nolinenumbers

\section{Introduction}
\label{sec:intro}

Present-day chemical abundance patterns offer crucial insights into the past evolution of galaxies. For example, radial abundance gradients arise from the complex interaction of gas accretion, interstellar medium (ISM) enrichment, mixing, and feedback tied to stellar evolution. Understanding how these gradients evolve over time helps constrain the various physical processes governing galactic evolution. However, tracking the changes in abundance patterns within a galaxy is challenging because history becomes erased by ISM mixing. Nevertheless, thanks to the different measurements, the evolution of radial metallicity gradients in galaxies across a wide range of masses is now available to high redshift. 

The nearby massive galaxies universally show negative gradients~\citep{Vila-Costas1992, Zaritsky1994, Ryder1995, vanZee1998, Pilyugin2004, Pilyugin2006, Pilyugin2007, Pilyugin2014, Moustakas2010, Gusev2012, Sanchez2014, Bresolin2015}. This result agrees with models based on the standard inside-out scenario of disk formation, which predict a relatively quick self-enrichment with oxygen abundances and an almost universal negative metallicity gradient once this is normalized to the galaxy's optical size~\citep{Boissier1999, Boissier2000}. At higher redshifts though, metallicity gradients show considerable variation among galaxies~\citep{Cresci2010, Queyrel2012, Stott2014, Carton2018}, where a notable fraction of galaxies exhibit shallow or even positive gradients.

Stars that formed in different regions and epochs of a galaxy eventually mixed, and now reveal only the present-day abundance distribution. In this context, the Milky Way~(MW) provides a unique opportunity to unravel its past. Encoded in each star's chemical composition is information regarding the star's time and place of birth \citep{2002freeman-BH, Ness2019, 2022Ratcliffe}, allowing for the detection and classification of accretion events \citep[e.g.][]{Helmi2018_gse, Buder2022, Cunningham2022, Khoperskov2023_hestiaHalo, Horta2023, Buder2024}, as well as an investigation into the chemical evolution \citep[e.g.][]{Matteucci2012, Minchev2018_rbirth, Frankel2018, Lu2022_Rb, Prantzos2023, 2020Ratcliffe, Ratcliffe2023_enrichment, Ratcliffe2023_chemicalclocks, Spitoni2024} and present-day state \citep[e.g.][]{hayden2015chemical, Ratcliffe2023_conditional, Hawkins2023, Imig2023, Haywood2024, Khoperskov2024} of the MW disk. 

While abundances contain a wealth of information on their own, we also need to know when and where the stars were born in order to best understand the evolutionary history of the MW disk as a whole. Since stars radially migrate away from their birth sites due to interactions with transient spirals (e.g., \citealt{Selwood2002, Roskar2008_migration, 2009schonrichBinney}), the overlap of multiple pattern speeds \citep{Minchev2006, Minchev2010, Marques2024}, or a rapidly slowing down bar \citep{Halle2015, Khoperskov2020, Wozniak2020}, a given location in the Galaxy can be comprised of stars with a variety of different birth radii depending on stellar age and metallicity \citep{Minchev2018_rbirth, Agertz2021_vintergatanI, Carrillo2023}. This causes mono-age abundance gradients measured in the present-day to appear weaker than the environment the stars formed in \citep{Pilkington2012, Minchev2013, Kubryk2013, Renaud2024}, and therefore radial migration must be taken into account when modeling the Galaxy's enrichment \citep{Francois1993}. Unfortunately, a star's angular momentum can permanently change, making it impractical to trace the star's orbit back to recover where it was born. 

\begin{figure*}
     \centering
     \includegraphics[width=.95\textwidth]{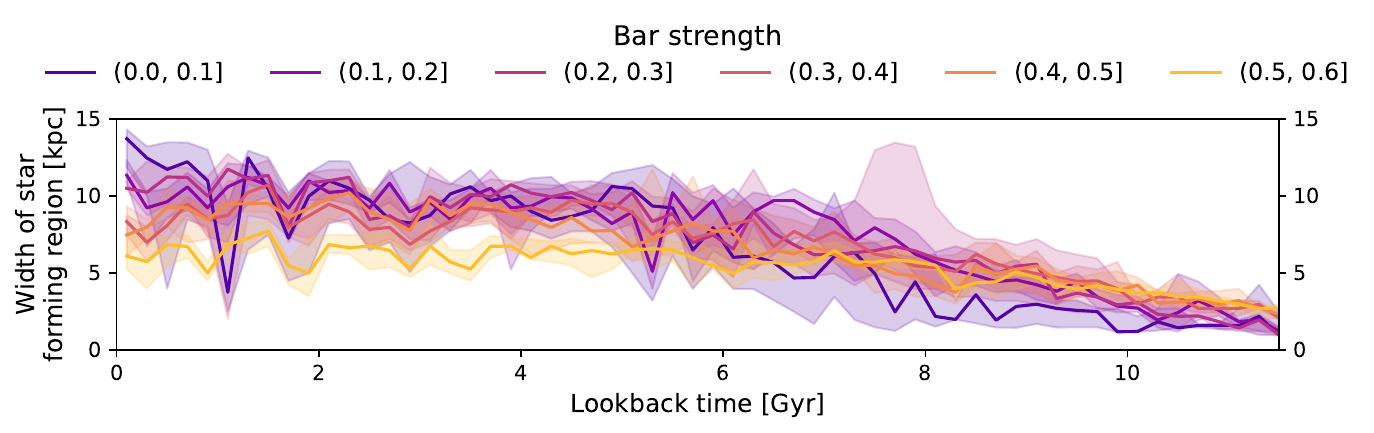}\\
\caption{Illustration of how the median width of the star-forming region (and 25-75\%-tile shaded region) in galaxies grows with cosmic time, as a function of present-day bar strength using TNG50 MW/M31-like galaxies. In weaker-barred galaxies, where the width of the star-forming region increases by over 10 kpc, the method of recovering the metallicity gradient from the range in [Fe/H] across age bins fails. When the width of the star-forming region does not change significantly over time (such as in the strongest barred galaxies), the method produces less error. Recreation of the top panel of Figure 6 from \cite{Ratcliffe2024_tng50}. } 
\label{fig:dr}
\end{figure*}

Recent works have thus leveraged the chemical homogeneity of stellar clusters \citep{BH2010, ness2022} to recover birth radii of stars in the MW disk \citep[e.g.][]{Minchev2018_rbirth, Frankel2018, 2019Frankel, Frankel2020}. As shown in observations \citep{Ness2019, Sharma2022} and cosmological zoom-in simulations \citep{Carrillo2023}, stars in small bins of [Fe/H] and age have small scatter in other abundances, indicating that these two variables alone can capture the primary evolution of the birth environment of the Galaxy. Leveraging this along with a linear birth metallicity gradient (which is found in observations \citealt{Deharveng2000, Esteban2017, ArellanoCordova2021}, and simulations \citealt{Vincenzo2018, Lu2022_sims}), stellar birth radii ($\rm \text{R}_\text{birth}$) can be recovered if the time evolution of the metallicity gradient ($\rm \nabla [Fe/H](\tau)$) and projected central metallicity ($\rm [Fe/H](R = 0, \tau$)) are known \citep{Kordopatis2015}. \cite{Minchev2018_rbirth} constructed a method which simultaneously recovered $\rm \text{R}_\text{birth}$ and the MW disk metallicity profile over time under the constraint that the resulting $\rm \text{R}_\text{birth}$ distributions should be physically meaningful. In discovering a linear relation between the scatter in [Fe/H] across age bins and the birth metallicity gradient evolution with cosmic time, \cite{Lu2022_Rb} was able to update this method so that recovering $\rm \text{R}_\text{birth}$ is fully self-consistent. 

In the realm of TNG50 MW/M31 analogues, this method was found to work well in galaxies with stronger bars \citep{Ratcliffe2024_tng50}. For galaxies with weaker bars, however, the metallicity gradient was typically unable to be recovered from the range in [Fe/H]. The ability to recover $\rm \nabla [Fe/H](\tau)$ from the scatter in [Fe/H] across age was found to depend on how the width of the star-forming region evolves with time; stronger barred galaxies tend to have only a minor increase in the region of where stars are forming in their disk, while in weaker barred galaxies it can grow drastically \citep{Ratcliffe2024_tng50}. Thus, to best capture the evolution of the metallicity profile, the growth of the disk needs to be accounted for \citep[][]{Molla2019}. 

In this paper, we propose a correction in recovering the metallicity gradient that accounts for how the width of the star-forming region varies with time as a function of present-day bar strength. This theoretical advancement enables us not only to determine the birth radii of Milky Way stars and constrain the amplitude of radial migration as a function of time and Galactocentric distance, but also to trace the evolution of the radial metallicity gradient back to a redshift of $\approx 3$. This approach bridges Galactic archaeology with nearby extragalactic and high-redshift observations, providing a more comprehensive view of the evolution of galaxies in general.

The paper is organized as follows. Sections \ref{sec:data} and \ref{sec:methods} describe the data and methods used in this work, with the correction described in \ref{section:method_correction}. The results of applying our correction to both simulations and observational data are given in Section \ref{sec:results}. Section \ref{sec::gradients_in_time} puts our MW results into context, and provides criteria to identify MW progenitors. We end with our conclusions in Section \ref{sec:conclusions}.

\section{Data}
\label{sec:data}

We used data from two simulation suites --- TNG50 and NIHAO-UHD --- in addition to APOGEE DR17 MW red giant disk stars. The MW/M31-like galaxies from TNG50 are used to measure how much the width of the star-forming region grows over time as a function of present-day bar strength. Two galaxies from NIHAO-UHD are used to show the success of the correction proposed in this paper to account for galactic growth when recovering the metallicity gradient from the scatter in [Fe/H] across age bins. Finally, we applied our correction to recover new estimates of stellar birth radii in the MW disk, which now better follow expectations of inside-out formation.

\subsection{TNG50 MW/M31 analogues}
\label{sec:data_tng50}

We analyzed MW/M31 analogues from the TNG50 cosmological simulation \citep{Pillepich2019, Nelson2019, Nelson2019a_TNG}, which is the highest resolution run of the IllustrisTNG project \citep{Pillepich2018_TNG, Marinacci2018_TNG, Naiman2018_TNG, Springel2018_TNG, Nelson2018_TNG, Nelson2019a_TNG} run with the moving-mesh
code AREPO \citep{Springel2010}. Within TNG50 there are 198 MW/M31-like galaxies, whose data are publicly available\footnote{https://www.tng-project.org/data/milkyway+andromeda/}. These galaxies were chosen using the criteria presented in \cite{Pillepich2023_MW_Sample} at redshift 0, which ensures that the galaxies (I) have disk-like morphology with spiral arms, (II) have a stellar mass in the range $M_*(< 30$kpc) = $10^{10.5-11.2}$M$_\odot$, (III) have no other galaxy with stellar mass $\geq 10^{10.5}$M$_\odot$ within 500 kpc, and (IV) have a halo total mass smaller than $M_{200c}$(host)$<10^{13}$M$_\odot$. We used the redshift 0 bar strengths from \cite{Khoperskov2023}, which are peak values of $m$ = 2 Fourier harmonics of the density distributions, and the disk scale lengths from \cite{SotilloRamos2022}, which are measured by fitting an exponential profile to the radial stellar surface density distribution excluding the bulge region.

To make our results derived in TNG50 applicable to the MW, we scaled each galaxy to have a disk scale length of 3.5 kpc and a rotational velocity of 238 km/s at 8 kpc \citep{BH2016}. We only considered the stellar disk, defined as $\rm \text{R}_\text{birth} < 15$ kpc, R < 15 kpc, |z| < 1 kpc, $|\rm \text{z}_\text{birth}| < $ 1
kpc, ecc < 0.5. We also only used stellar particles that were born and are currently bound to the galaxy. Following \cite{Ratcliffe2024_tng50}, we only examine the 178 galaxies that have a strong correlation ($< - 0.85$) between [Fe/H] and $\rm R_{birth}$ in the most recent Gyrs of evolution. 

\subsection{NIHAO-UHD simulations}
\label{sec:data_buck}

We used the galaxies g2.79e12 and g8.26e11 from the NIHAO-UHD simulation suite \citep{Buck2019, 2020buck_NIHAO-UHD}, run using the smooth-particle hydrodynamics code GASOLINE2 \citep{2017gasoline2}. Both galaxies have MW-like mass, a boxy/peanut-shaped bulge, and have been studied extensively due to their similarity to the MW in terms of chemistry \citep{2020_buckchemical, 2021BuckHD_chemEnrich,2022Ratcliffe,Wang2023}, merger history \citep{Buck2023}, dark halo spin \citep{Obreja2022} and satellite galaxy properties \citep{Buck2019}. g2.79.e12 is the most MW-like in terms of its bar and spiral arms, whose properties have been studied and put in the context of the MW \citep{Buck2018, 2019ApJ...874...67B, Hilmi2020, Vislosky2024, Marques2024}. The bar strengths were estimated from the ratio of the amplitude of the $m=2$ to $m=0$ Fourier components of the stellar density, and found to be  $A_2/A_0\approx0.4 - 0.5$ for g2.79e12 (Figure 1 of \citealt{Hilmi2020}) and $< 0.1$ for g8.26e11.

Similar to our TNG50 sample, we scaled the NIHAO-UHD galaxies to have a disk radial scale length of 3.5 kpc, and chose stellar disk particles by $\rm \text{R}_\text{birth} < 15$ kpc, |z| < 1 kpc, $|\rm \text{z}_\text{birth}| < $ 1 kpc, and ecc < 0.5. We additionally selected stellar particles with $|\rm \text{V}_\text{z,birth}| < $ 50 km/s to minimize merger contamination. To best match our MW APOGEE sample, we used star particles between 4 < R < 12 kpc, and only considered a random sample of 160,000 stellar particles from each galaxy.

\begin{figure}
     \centering
     \includegraphics[width=.5\textwidth]{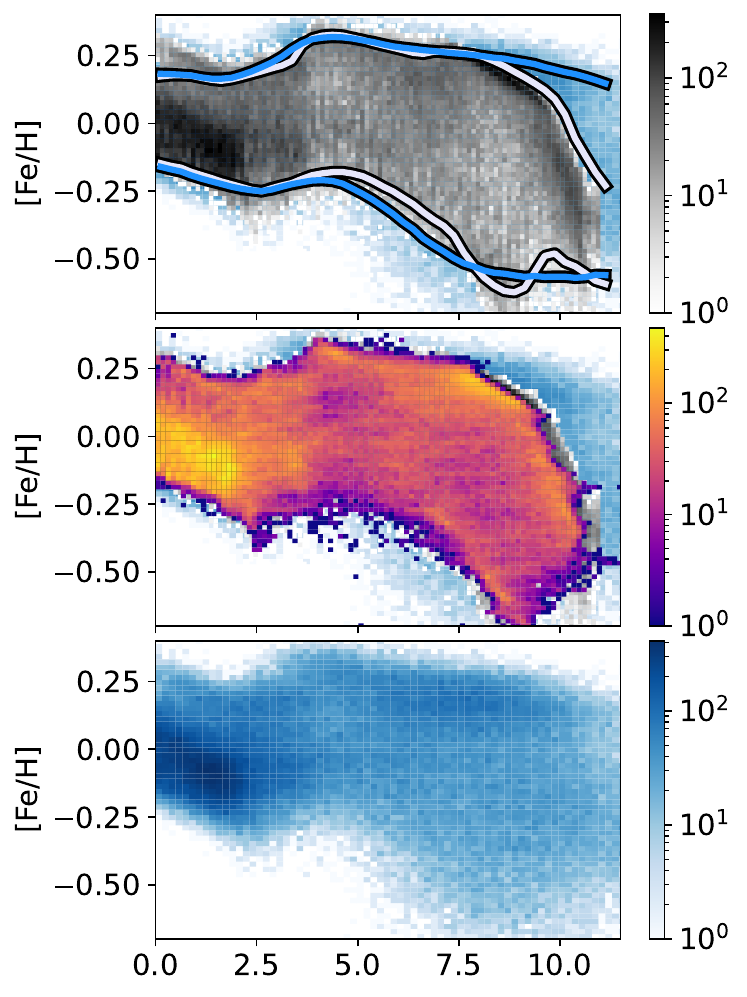}\\
\caption{Role of denoising the age--metallicity relation in recovering the metallicity gradient from the scatter in [Fe/H] as a function of age. {\bf Top:} The age--metallicity relation for g2.79e12 (grey scale) with 15\% age and 0.05 dex [Fe/H] uncertainty added (blue scale). The horizontal lines correspond to the upper and lower bounds used to calculate Range[Fe/H](age) measured with (blue) and without (white) measurement uncertainties. The range measured using the data with uncertainties is much larger than the range measured using the true age and [Fe/H] values for age $>9$ Gyr. {\bf Middle:} The distribution of the denoised ages (Section \ref{method:denoise}) and [Fe/H] (plasma) match well the age--metallicity relation of the true distribution. The presence of present-day observational uncertainties (also shown in the {\bf bottom} panel) makes denoising the age--metallicity relation a requirement to recover the structure in this plane, and thus Range[Fe/H](age). } 
\label{fig:buck_denoise}
\end{figure}

\subsection{APOGEE DR17}
\label{sec:data_apogee}

To recover the metallicity profile with cosmic time and radius for the MW disk (and therefore, also stellar $\rm \text{R}_\text{birth}$), we used data from the Sloan Digital Sky Survey IV (SDSS-IV; \citealt{blanton2017sloan}) APOGEE DR17 \citep{Majewski2017, apogeeDR17} catalog. Stellar parameters and abundances are processed by the APOGEE Stellar Parameter and
Chemical Abundance Pipeline (ASPCAP; \citealt{Holtzman2015, GarciaPerez2016aspcap, Jonsson2020}). We additionally used orbital eccentricities from the astroNN catalog \citep{Mackereth2018, LeungBovy2019} and the recommended uncorrected Sharma model ages from the \texttt{distmass} catalog \citep{Imig2023, StoneMartinez2024}. We selected disk ($|\text{z}| < 1$ kpc, ecc $< 0.5$, |[Fe/H]| $< 1$) stars with small uncertainties ([Fe/H]$_\text{err} < 0.05$ dex, unflagged [Fe/H], $\rm log_{10}Age_\text{err; upper} > 0$), and used red giant branch stars to avoid systematic abundance trends ($2 < \text{logg} < 3.6, 4250 < \text{T}_\text{eff} < 5500$ K). We also ensured to use only stars with parameters that are within the \texttt{distmass} training set by removing stars with bit 2 set. These cuts left us with 162,784 MW disk stars for our sample.

In order to illustrate the direct effect of the correction on the gradient, we also used the sample of red giant APOGEE DR17 disk stars from \cite{Ratcliffe2023_enrichment}. This sample has the same selection criteria as listed above, except it uses the age catalog of \cite{Anders2023_ages}.

\section{Methods}\label{sec:methods}

\subsection{Proposed correction}\label{section:method_correction}

The metallicity gradient at a given lookback time ($ \rm \nabla [Fe/H](\tau)$) can be written as 
\begin{equation}\label{eqn:grad}
    \rm \nabla [Fe/H](\tau) = \frac{\Delta\text{[Fe/H]}(\tau)}{\Delta\text{R}(\tau)},
\end{equation}
where $\rm \Delta\text{[Fe/H]}(\tau)$ is the difference between the maximum and minimum [Fe/H] at a given time $\tau$, and $\rm \Delta\text{R}(\tau)$ is the width of the star-forming region at a given time $\tau$. 
We are unable to accurately estimate $\Delta\text{[Fe/H]}(\tau)$ due to observational constraints, such as incompleteness and pipeline dependencies. However, the relative scatter in [Fe/H] (Range[Fe/H]) can be estimated from the $5-95\%$-tile range \citep{Lu2022_Rb, Ratcliffe2024_tng50} or the standard deviation in [Fe/H] \citep{Ratcliffe2023_enrichment} in different age bins. Both of these methods minimize contamination from outliers and capture the shape of $\Delta\text{[Fe/H]}(\tau)$.

To recover $\rm \nabla [Fe/H](\tau)$, the method of \cite{Lu2022_Rb} assumes that the width of the star-forming region is constant with time, or $\Delta\text{R}(\tau)$ = $\Delta\text{R}$. However, Figure \ref{fig:dr} clearly illustrates that $\Delta\text{R}(\tau)$ grows with time. Since each group of galaxies based on redshift 0 bar strength exhibits a distinct trend in an increasing $\Delta\text{R}(\tau)$, the method to recover the metallicity gradient from the range in [Fe/H] needs to allow for this variation. We therefore propose using a smoothed version of $\Delta\text{R}(\tau)$ from Figure \ref{fig:dr} to correct for the non-constant width of the star-forming region. Smoothing the width of the star-forming region for each bar strength (measured every 0.2 Gyr) over 1.2 Gyr captures the overall shape of $\rm \Delta R(\tau)$ while minimizing the local variations that may be non-universal (see Figure \ref{fig:correction} in the Appendix). Our results below are consistent for different smoothing bins.

Since the metallicity gradient and scatter are negatively correlated, this means that the normalized gradient and scatter are additive inverses; when the scatter is largest, the metallicity gradient is steepest (i.e. the most negative). We introduce the corrected metallicity scatter as 
$$\rm \sigma[Fe/H](age) \equiv\frac{\text{Range[Fe/H]}(age)}{\Delta\text{R}(age)},$$ 
and then the normalized scatter  can be written as: 
$$\rm \widetilde{\sigma}[Fe/H](age) \equiv \rm \frac{\sigma[Fe/H](age) - min(\sigma[Fe/H](age))}{\max(\sigma[Fe/H](age)) - min(\sigma[Fe/H](age))}.$$
Given the gradient is negative, we need 0 to represent the weakest gradient, and -1 to represent when the gradient is steepest. The normalization can be rewritten as follows:
\begin{align*}
    \rm \widetilde{\nabla}[Fe/H](\tau) 
 & \equiv \rm \frac{\nabla[Fe/H](\tau) - min(\nabla[Fe/H](\tau))}{\max(\nabla[Fe/H](\tau)) - min(\nabla[Fe/H](\tau))} -1\\
 & \equiv \rm \frac{\nabla[Fe/H](\tau) - max(\nabla[Fe/H](\tau))}{\max(\nabla[Fe/H](\tau)) - min(\nabla[Fe/H](\tau))}.
\end{align*}
As $\rm \widetilde{\nabla}[Fe/H](\tau)$ and $\rm \widetilde{\sigma}[Fe/H](age)$ are additive inverses, we then have
$$\rm\widetilde{\sigma}[Fe/H](age) = - \widetilde{\nabla}[Fe/H](\tau),$$
or 
\begin{align*}
    \rm\widetilde{\sigma}[Fe/H](age) &= \rm-\frac{\nabla[Fe/H](\tau) - max(\nabla[Fe/H](\tau))}{\max(\nabla[Fe/H](\tau)) - min(\nabla[Fe/H](\tau))}\\
    & =\rm \frac{\nabla[Fe/H](\tau) - max(\nabla[Fe/H](\tau))}{\min(\nabla[Fe/H](\tau)) - max(\nabla[Fe/H](\tau))}.
\end{align*}
Following \cite{Lu2022_Rb}, we define 
$$\rm b \equiv max(\nabla [Fe/H](\tau))$$ 
$$\rm a \equiv min(\nabla [Fe/H](\tau)) - max(\nabla [Fe/H](\tau)),$$
which allows us to rearrange the above equation to solve for $\rm\nabla[Fe/H](\tau)$:
\begin{equation}\label{eqn:solved}
    \rm \nabla [Fe/H](\tau) = a\,\widetilde{\sigma}[Fe/H](age) + b.
\end{equation}
Thus, we have that the time evolution of the metallicity gradient is a linear function of the scatter in [Fe/H] across age after adjusting it for the growth of the star-forming region.

The MW disk birth gradient is expected to weaken with cosmic time \citep[e.g.][]{Spitoni2023_spirals, Prantzos2023}, and therefore, the gradient of the youngest stars in the MW can be used for variable b in Equation \ref{eqn:solved}. The steepest gradient (a + b in Equation \ref{eqn:solved}), on the other hand, is unknown, though it can be visually inferred from the distribution of birth radii for solar neighborhood populations (\citealt{Lu2022_Rb, Ratcliffe2023_enrichment}). In this work, we automate the determination of the steepest gradient, rather than the manual by-eye approach used in previous works. We find the optimal value by minimizing $\rm |R - R_{birth}|$ for the youngest ($<2$ Gyr) stars across the entire disk, where $\rm R_{birth}$ is derived following Section 3 from \cite{Ratcliffe2023_enrichment}: $$\rm R_{birth} = \frac{[Fe/H] - [Fe/H](R = 0, \tau)}{a\,\widetilde{\sigma}[Fe/H](age) + b}.$$ The optimization is achieved by minimizing the objective function
$$
  \rm \min_{a < 0} \left\|R - \frac{[Fe/H] - [Fe/H](R = 0, \tau)}{a\,\widetilde{\sigma}[Fe/H](age) + b} \right\|_1
$$

We solve this minimization problem using the L-BFGS-B algorithm from scipy.optimize \citep{Byrd1995_LBFGS}, its default solver for bound-constrained minimization. We find the mean optimal steepest gradient is -0.136 dex/kpc across 100 Monte Carlo samples, which is slightly weaker than previously reported (see Section \ref{sec:results_mw} for more discussion).

\subsection{Denoising the age--metallicity relation}\label{method:denoise}


In this section, we show that denoising the age--metallicity plane before calculating Range[Fe/H]($\tau$) is a necessary step in the realm of present-day observational uncertainties in age. To make the best comparisons with the MW disk, we added error in age (15\%) and [Fe/H] (0.05 dex) that is in line with observational uncertainties of current large spectroscopic surveys \citep[e.g.][]{Jonsson2020, Anders2023_ages}. The top panel of Figure \ref{fig:buck_denoise} illustrates the structure in the age--metallicity plane of g2.79e12 that is lost due to observational uncertainty (see also \citealt{Renaud2021_vintergatanII}). In particular, the range in [Fe/H] is significantly larger at older ages for the data with uncertainties (blue) than the true data (grey). This increase in the [Fe/H] scatter is due to the larger age error for older age stars, along with the sharp change in [Fe/H] for increasing lookback time at the earlier stages of the disk. Therefore, this significant overestimation of Range[Fe/H] for age $> 8$ Gyr is not found for the younger ages, and causes the recovered $\rm \nabla [Fe/H](\tau)$ to be too steep at larger lookback times, an issue not found when using the unperturbed data (see Appendix Figure \ref{fig:buck_grads}). 

To tackle this problem caused by observational uncertainties, we denoised the age--metallicity relation using the nonparametric maximum likelihood estimator (NPMLE; \citealt{Soloff2021_npmle})\footnote{https://github.com/jake-soloff/NPEB/tree/master}. The NPMLE recovers denoised estimates of parameters without assuming the data follow a specific underlying distribution. This allows for better estimates of complex data without simplifying the naturally intricate relationship between variables. The main details of the method are given in Appendix Section \ref{sec:methods_npmle}. The denoised age--metallicity relation of g2.79e12 matches well the true distribution (middle panel of Figure \ref{fig:buck_denoise}). In particular, the shape in [Fe/H] for ages $> 8$ Gyr is in much better agreement with the true data than the error-convolved data. 

\section{Recovering the ISM metallicity gradient from stellar age and [Fe/H]}
\label{sec:results}

\begin{figure}
     \centering
     \includegraphics[width=.45\textwidth]{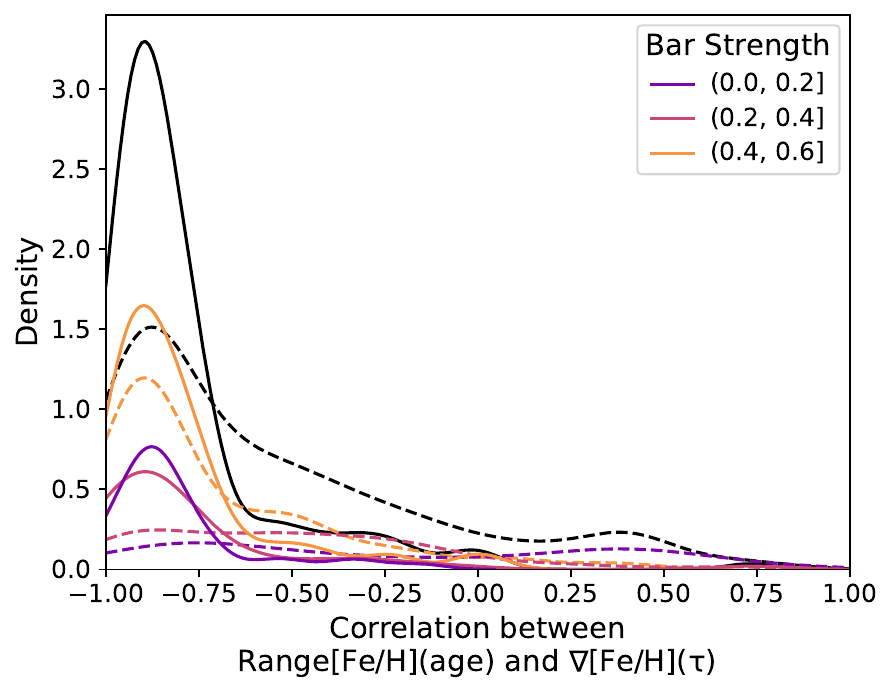}\\
\caption{Correlation between the shape of the metallicity gradient over time and the range in [Fe/H] measured across age bins using the method with (solid lines) and without (dashed lines) our proposed correction applied for our sample of 178 TNG50 MW/M31-like galaxies. The colored lines correspond to the correlation of galaxies with different bar strengths, with the black line providing the density of the entire sample. A correlation of -1 implies that the shape of the gradient can be perfectly recovered from Range[Fe/H](age), and that the method is successful. With no correction applied, the median correlation is -0.72 (median absolute percent error of 31\%), whereas when applying the correction the median correlation improves to -0.86 (median absolute percent error of 21\%). The correction has a smaller improvement on stronger-barred galaxies (yellow) where the method already exhibited good success, but the correlation of galaxies with weaker present-day bars show an improvement of over 75\%. } 
\label{fig:density}
\end{figure}

\subsection{Correcting for the growth of the star-forming region in simulations}\label{sec:results_correction}

\subsubsection{Impact of the star-forming region growth correction in TNG50}

As discussed in Section \ref{section:method_correction}, we propose a correction to the measured scatter in [Fe/H] across age that takes the growth of the star-forming region into account by adjusting for $\Delta\text{R}(\tau)$. The time evolution of $\Delta\text{R}(\tau)$ depends on the galaxy's present-day bar strength; as a galaxy grows, quenching in the inner region of strongly barred galaxies \citep{Khoperskov2018, Geron2024} creates a more constant $\Delta\text{R}(\tau)$ with time, while weakly barred galaxies do not experience this halt in star formation, causing the width of the star-forming region to continually grow \citep{Ratcliffe2024_tng50}. Therefore, the $\rm \Delta R(\tau)$ correction is chosen based on the galaxy's redshift 0 bar strength (Figure \ref{fig:dr}, see also Figure \ref{fig:correction} in the Appendix).

To illustrate the power of our proposed correction, we first applied it to our TNG50 MW/M31 analogues sample. Figure \ref{fig:density} shows the correlation between $\rm \nabla [Fe/H](\tau)$ and Range[Fe/H](age) using the measured range (dashed lines) and adjusted range using our proposed correction (solid lines). A correlation close to -1 implies recovering the shape of $\rm \nabla [Fe/H](\tau)$ from Range[Fe/H](age) is successful, whereas a correlation near 0 or positive implies that the method would be unsuccessful. Each group of galaxies exhibits an overall improvement in their ability to have the gradient recovered from the scatter in [Fe/H] after our proposed adjustment, with the median correlation across all galaxies improving to -0.86. In particular, the galaxies with bar strength < 0.4 show the largest improvement after applying our correction, with a peak in correlation now near -0.86. Our correction to the method only minimally helped in stronger barred galaxies, as the method was already quite successful in this regime. The overall enhancement our correction produces in the method is also reflected in the median absolute percent error when estimating $\rm \nabla [Fe/H](\tau)$ from Range[Fe/H](age), which decreases from 31\% to 21\% across all galaxies.

\begin{figure*}
     \centering
     \includegraphics[width=.465\textwidth]{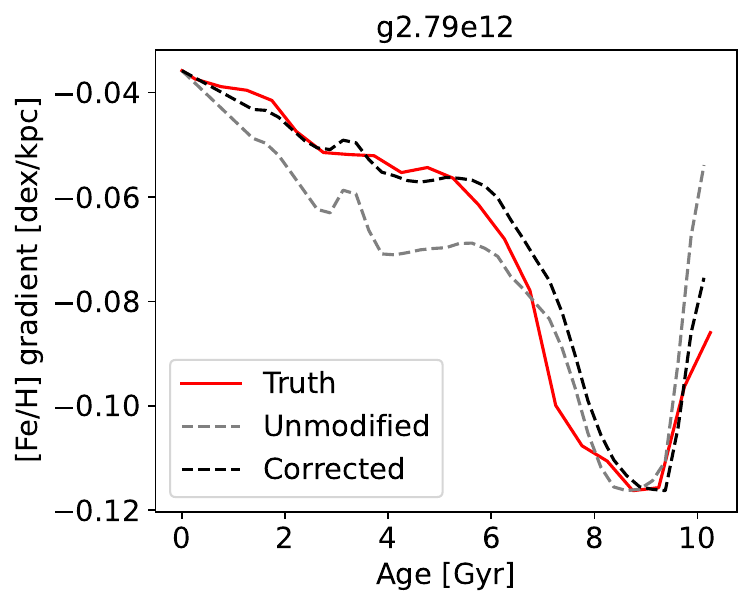}
     \includegraphics[width=.485\textwidth]{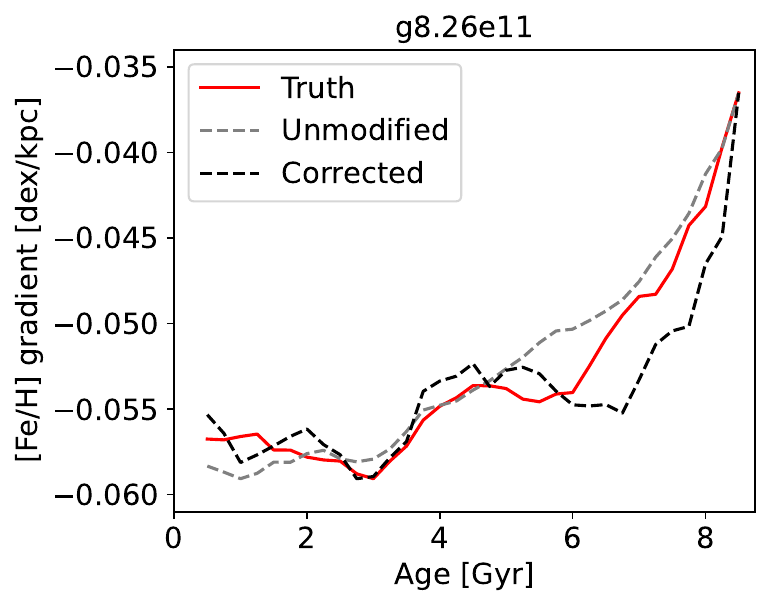}\\
\caption{Demonstration of how the recovered [Fe/H] gradient using our proposed correction works in {\bf left:} g2.79e12 (strong bar) and {\bf right:} g8.26e11 (no bar) from the NIHAO-UHD simulation suite, which are completely independent of the TNG50 galaxies which are used to determine the correction. As discussed in Section \ref{sec:results_buck}, we add 0.05 dex and 15\% error to [Fe/H] and age, and then denoise to make the best extensions to the MW. The black line in each panel represents the recovered metallicity gradient with the correction applied, and better matches the true [Fe/H] gradient evolution of disk stars (red line) than recovering the metallicity gradient without the correction (grey line). } 
\label{fig:buck}
\end{figure*}

\subsubsection{Confirming success with an independent simulation prescription --- NIHAO-UHD}
\label{sec:results_buck}

To demonstrate the correction we propose --- which is based solely on TNG50 MW/M31 analogues --- is not simulation-specific, we also show its success when recovering the gradient of MW-like galaxy g2.79e12 from NIHAO-UHD \citep{2020buck_NIHAO-UHD}. To ensure the correction can be extended to the MW, we added uncertainties in age (15\%) and [Fe/H] (0.05 dex). Since the range in [Fe/H] is overestimated at larger ages due to the uncertainties added (Section \ref{method:denoise}), we show the recovered gradients after denoising in age and [Fe/H] in Figure \ref{fig:buck}. Using the unmodified Range[Fe/H](age) (grey line, as proposed in \citealt{Lu2022_Rb}) does a reasonable job of recovering the true evolution of the metallicity gradient (red line), however it significantly underestimates the gradient at older ages. This is because stars are forming on a smaller area at these higher lookback times than compared to the later stages of evolution (Figure \ref{fig:dr}), which causes the [Fe/H] range to be smaller than if stars were forming on a similar radial width of the disk over time. Similarly, the range for younger stars is artificially larger due to stars forming on a larger range of the disk. Applying our $\rm \Delta R(\tau)$ correction to adjust the range in [Fe/H] and account for this disk growth allows for a closer recovery of the truth in the stronger-barred galaxy (black line in the left panel of Figure \ref{fig:buck}). We find that the median absolute percent error after accounting for the time variation of the width of the star-forming region is over three times smaller than when no correction is applied.

In addition to a NIHAO-UHD galaxy with a stronger bar, we also tested our correction for galaxies with no bar. Figure \ref{fig:dr} shows that the weaker-barred galaxies have a lot more variability in the growth of their star-forming region, particularly for galaxies with a bar strength $<0.1$. Our sample size for this regime of bar strength is also much smaller than the other groups. Thus, we combine bar strengths $0-0.2$ into one group for our correction. The right panel of Figure \ref{fig:buck} shows the recovered gradient for g8.26e11 with (black line) and without (grey line) correcting for the growth of the galactic disk. The correction gives an overall improvement to the recovered gradient, especially capturing the non-monotonic behavior. Although, there are additional, albeit minor, peaks created throughout that do not appear in the correction for galaxies with stronger bars. These minor fluctuations are due to the larger variation weaker barred galaxies exhibit in their growth (Figure \ref{fig:dr}), causing the growth correction to be less smooth. We also find that the gradient is estimated to be steeper than expected for $6-8$ Gyr, indicating that the correction for weaker barred galaxies is less universal than suggested using our TNG50 sample (Figure \ref{fig:density}). However, adjusting for $\rm \Delta R$ in this regime can still provide helpful corrections to the gradient.

\subsection{Recovering the evolution of the MW disk metallicity gradient in the ISM}
\label{sec:results_mw}

\begin{figure}
     \centering
     \includegraphics[width=.475\textwidth]{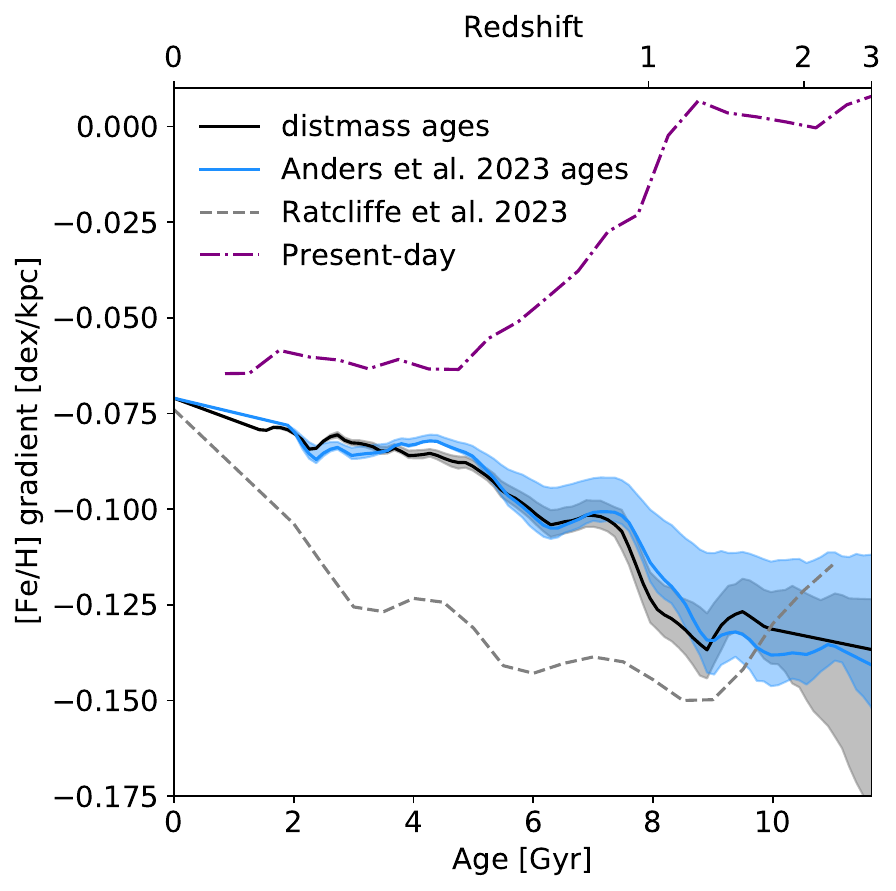}\\
\caption{Comparing recovered MW disk metallicity gradients with and without correcting for the growth of the star-forming region. The grey dashed line represents the gradient found in \cite{Ratcliffe2023_enrichment} using ages from \cite{Anders2023_ages}, while $\rm \nabla [Fe/H](\tau)$ recovered using the correction proposed in this paper is shown in blue (ages from \citealt{Anders2023_ages}) and grey (ages from \texttt{distmass}). The gradient is taken as the mean of 100 Monte Carlo samples where (age, [Fe/H]) are redrawn from a normal distribution and then denoised, and the shaded areas represent the 25 - 75\%-tiles. The range in [Fe/H] is measured in age bin widths of 1.5 Gyr, taken every 0.2 Gyr. The corrected gradient recovered using the two different age catalogs is strikingly consistent and illustrates that the fluctuations in the gradient are not artifacts of the age catalog. The steepening events in the gradient $\sim 8$ Gyr ago and $\sim 6$ Gyr ago are less significant after the correction is applied, however there is still non-monotonicity in the gradient's evolution. We also provide the present-day gradient (purple dashed line measured using $\rm R_{guide}$) to illustrate the information lost due to radial migration.} 
\label{fig:mw_grad}
\end{figure}

Now that we have established our proposed $\rm \Delta R (\tau)$ correction allows for a better recovery of $\rm \nabla [Fe/H](\tau)$, we extend it to the MW disk. We chose to apply the correction for galaxies with a bar strength of $0.5-0.6$, as the MW closely resembles this category in the realm of TNG50 MW/M31-like galaxies, including the present-day stellar mass and star formation history \citep{Khoperskov2023}. Figure \ref{fig:mw_grad} compares the metallicity gradient recovered after applying the correction to our APOGEE samples with ages from \texttt{distmass} (solid black line with the shaded area showing the 25-75\%-tile) and \cite{Anders2023_ages} (blue line) to that of \cite{Ratcliffe2023_enrichment} (dashed line) where no correction was applied. The gradients in this work are taken as the mean $\rm \nabla [Fe/H](\tau)$ of 100 Monte Carlo samples, where both the age and [Fe/H] were drawn from a normal distribution about their measured values and a standard deviation of their measurement errors and then denoised. Since stars older than 10 Gyr are not well-represented in the \texttt{distmass} catalog \citep{Imig2023}, we estimate the gradient using a linear regression fit from the $8-10$ Gyr data. The gradient recovered from \texttt{distmass} ages, including the estimation at older ages, is in good agreement with the gradient found using ages from \cite{Anders2023_ages} after the correction is applied, with both gradients showing similar trends across time. 

The overall trend of the newly recovered metallicity gradients is similar to what was previously reported in \cite{Ratcliffe2023_enrichment}; the gradient was steepest at $\sim 9-10$ Gyr ago and has been non-monotonically weakening with time. However, due to the correction, we now find a much weaker gradient in the past $\sim8$ Gyr, and a stronger gradient at older ages. The steepening phases found  $\sim 9$ Gyr ago potentially due to Gaia-Sausage-Enceladus (GSE; \citealt{Belokurov2018,Helmi2018_gse}) and 6 Gyr ago due to the Sagittarius dwarf galaxy \citep{Ibata1994, Law2010} are still found, and suggest that these steepening events may be real as they are found across different surveys, stellar evolutionary phases, and age catalogs. One key difference, though, is that the gradient at $\sim 6$ Gyr ago is weaker than when no correction was applied~($-0.1$ dex/kpc versus $-0.14$ dex/kpc). Additionally, the steepening phase at $\sim 4$ Gyr ago found in \cite{Ratcliffe2023_enrichment} using the \cite{Anders2023_ages} (APOGEE DR17 red giant disk stars) and {\it StarHorse} (GALAH DR3 subgiant branch disk stars; \citealt{Buder2021, Queiroz2023_SH}) age catalogs still exists after applying the correction to the \cite{Anders2023_ages} ages, however, it has a much smaller impact. The lack of fluctuation found at this time using the \texttt{distmass} catalog suggests that the exponential age errors may have washed away the small feature\footnote{The \texttt{distmass} catalog provides age errors in terms of $\rm \log_{10} yr$.}. A more detailed discussion of the MW disk ISM gradient evolution is given in Section \ref{sec::gradients_in_time_ISM}, where the gradient is discussed in a broader context.

Figure \ref{fig:mw_grad} also provides the present-day gradient measured using linear regression in the [Fe/H]--$\rm R_{guide}$ plane across different mono-age populations. The flat gradient of stars $\gtrapprox 8$ Gyr is when the correlation between [Fe/H] and $\rm R_{guide}$ significantly weakens, thus the gradient for these older ages is capturing a cloud-like structure rather than a true nearly flat gradient. Examining the present-day gradient, we find three main epochs: $0-5$ Gyr, $5-9$ Gyr, and $>9$ Gyr. In the most recent 5 Gyr of evolution, we find that radial migration affects the metallicity gradient by weakening it $\lesssim 0.03$ dex/kpc, suggesting that the MW disk has had quiescent evolution over the past $\sim 5$ Gyr. From $5-9$ Gyr ago, we find a significant weakening of the present-day stellar gradient, while for stars older than 9 Gyr (during the transition from the low- to high-$\alpha$ sequence) we find no [Fe/H]-$\rm R_{guide}$ structure. Comparing this to our recovered [Fe/H]--$\rm R_{birth}$ gradient (black and blue lines), we see that these epochs correspond with the steepening of the gradient potentially from Sagittarius and GSE~\citep{Annem2024,Buck2023}. 


\subsubsection{Improved stellar birth radii for the MW disk}

\begin{figure*}
    \centering
    \includegraphics[width=1\textwidth]{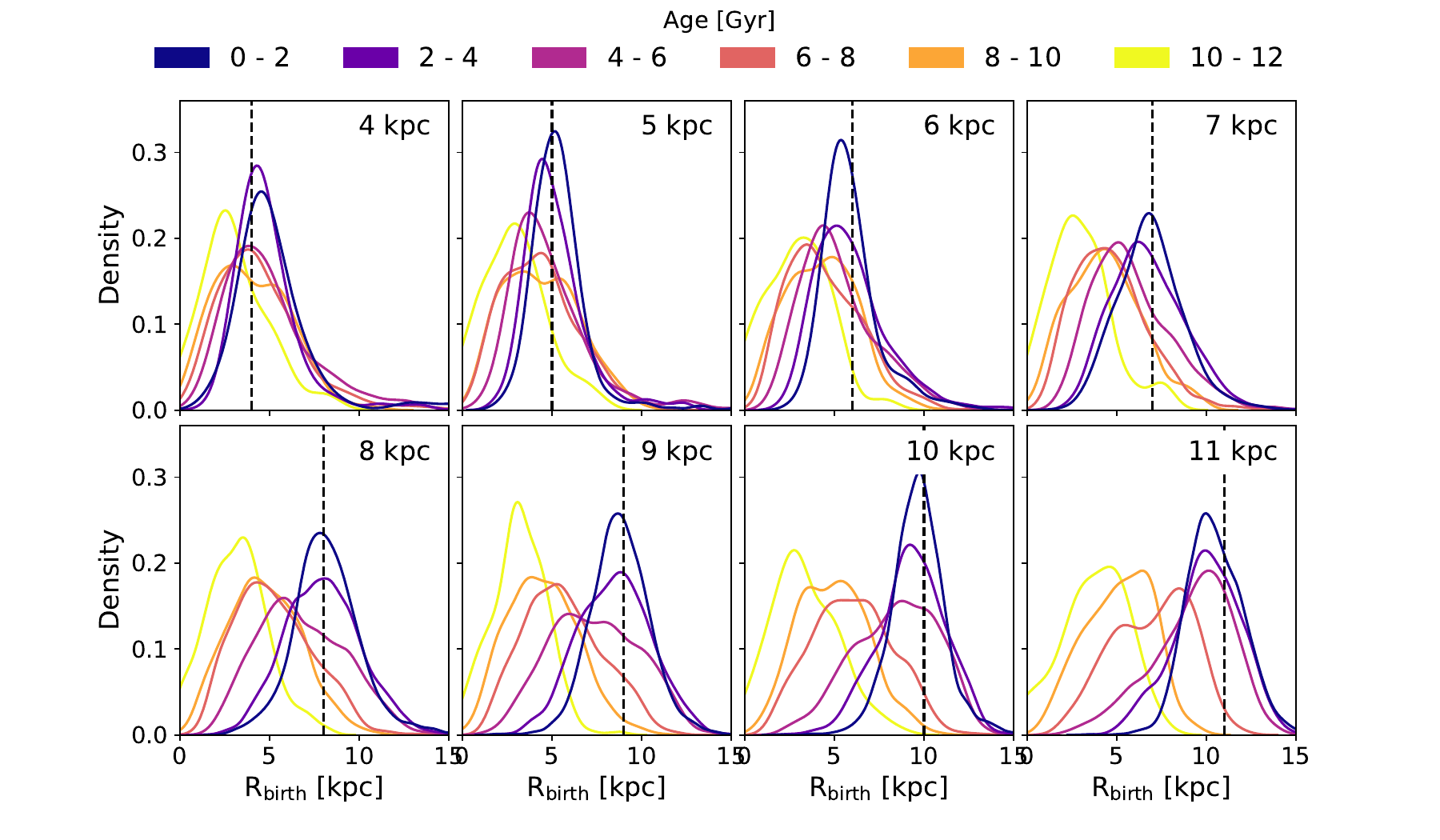}\\
\caption{Birth radii distributions of mono-age MW disk stars currently located within 0.25 dex of different radii (as indicated by the dashed black line in each panel) with $|z|\leq 0.5$ kpc. The area under each mono-age curve is normalized to 1, so the curves are not representative of the relative proportion of each age at a given radii, but rather to illustrate the individual trends of each mono-age population. For each radial bin, the youngest population peaks within 0.5 kpc from where they are currently located today while each older population has consecutively smaller and smaller birth radii. Comparing the distribution of stars in the solar neighborhood (bottom left panel) to that of \cite{Ratcliffe2023_enrichment}, the correction used in this paper provides less migration for the $2-6$ Gyr old populations and less spread in the oldest population. } 
\label{fig:SNdist}
\end{figure*}

With our corrected $\rm \nabla [Fe/H](\tau)$ (Figure \ref{fig:mw_grad}), we can recover improved stellar birth radii of the MW disk. Just as in \cite{Lu2022_Rb} and \cite{Ratcliffe2023_enrichment}, the central metallicity is taken as the 95\%-tile in [Fe/H] for older ages (age $>=10.75$ Gyr) and estimated as monotonically enriching until its projected value at the present-day assuming the youngest stars in the solar neighborhood have a metallicity of 0.06 \citep{Nieva2012, Asplund2021} and a gradient of -0.071 dex/kpc \citep{Trentin2024}. As discussed in \cite{Ratcliffe2024_tng50}, this assumption is valid, providing the $\rm R_\text{birth}$ distributions of mono-age populations in the solar neighborhood follow expectations of inside-out formation.

Figure \ref{fig:SNdist} shows the distribution of mono-age populations of stars currently located at different radii, as indicated by the black dashed line in each panel. Previously, only the distributions of mono-age populations in the solar neighborhood were used to confirm if the birth radii seemed reasonable. Here, we now ensure that the distributions of mono-age populations are reasonable for stars currently located at a variety of radii throughout the Galactic disk. At each radius, the youngest stars are born predominantly within 0.5 kpc of where they are currently located, and the older stars peak at progressively smaller birth radii in agreement with expectations from inside-out formation \citep[e.g.][]{Minchev2014, Agertz2021_vintergatanI, Ratcliffe2024_tng50}. These birth radii also show less migration than \cite{Lu2022_Rb, Ratcliffe2023_enrichment}, particularly for the middle-aged populations. The oldest population also exhibits less spread in birth radii than in \cite{Ratcliffe2023_enrichment}, though it does contain negative birth radii. One possibility for this is that these older stars are both not well represented in the data and also have the largest age uncertainties, and therefore, the upper envelope used to estimate the projected central metallicity at these older ages may not adequately capture the true distribution of the projected central metallicity. However, these negative birth radii make up less than $0.1$\% of our $\rm R_{birth}$ estimates, which is very minimal contamination. 

We can now also get an improved estimate of the birth radius of the Sun. Assuming $\rm [Fe/H]_{Sun} = 0 \pm 0.05$ dex \citep{Asplund2009} and $\rm age_{Sun} = 4.57 \pm 0.11$ Gyr \citep{Bonanno2002}, we estimate the Sun's birth radius as $6.66 \pm 0.58$ kpc. This result is in great agreement with that found by \cite{Wielen1996}, who estimated this value simply from the difference in [Fe/H] of the Sun versus other stars in the solar neighborhood and assuming a constant metallicity gradient with time. Our estimate is in the middle of other solar birth radius estimates, which range from 4.5 kpc to 7.8 kpc \citep{Kubryk2015, Frankel2018, Minchev2018_rbirth, Frankel2020, Lu2022_Rb, Baba2023}.

\subsubsection{Migration strength in the MW disk}\label{sec::migration_strength}

\begin{table*}[h]
    \centering
    \begin{tabular}{c c c c c c}
        \toprule
        & \multicolumn{5}{c}{Median $\pm$ Std migration distance [kpc] } \\
        \cmidrule(lr){2-6}
       Age [Gyr] & $\rm 3 < R_{guide} < 5$ & $\rm 5 < R_{guide} < 7$ & $\rm 7 < R_{guide} < 9$ & $\rm 9 < R_{guide} < 11$ & All $\rm  R_{guide}$\\
        \midrule
        $0-2$ & -1.0 $\pm$ 2.6 & 0.1 $\pm$ 2.5 & 0.0 $\pm$ 1.9  & 0.0 $\pm$ 1.6 & 0.0 $\pm$ 1.9\\
        $2-4$ & -0.8 $\pm$ 1.9 & 0.1 $\pm$ 2.5 & 0.1 $\pm$ 2.1  & 0.2 $\pm$ 1.9 & 0.1 $\pm$ 2.1\\
        $4-6$ & -0.6 $\pm$ 2.8 & 0.8 $\pm$ 2.5 & 1.1 $\pm$ 2.4  & 0.8 $\pm$ 2.2 & 0.9 $\pm$ 2.4\\
        $6-8$ & -0.5 $\pm$ 2.5 & 1.4 $\pm$ 2.3 & 2.6 $\pm$ 2.2  & 3.1 $\pm$ 2.2 & 2.1 $\pm$ 2.6\\
        $8-10$ & -1.0 $\pm$ 2.2 & 1.1 $\pm$ 2.2 & 3.3 $\pm$ 2.0  & 4.6 $\pm$ 1.9 & 2.2 $\pm$ 2.8\\
        $10-12$ & 0.6 $\pm$ 1.9 & 2.7 $\pm$ 1.8 & 4.7 $\pm$ 1.6 & 6.2 $\pm$ 1.6 & 3.6 $\pm$ 2.6\\
        All ages & -0.7 $\pm$ 2.5 & 0.9 $\pm$ 2.5 & 1.2 $\pm$ 2.5  & 0.7 $\pm$ 2.4 & 0.8 $\pm$ 2.6\\
        \bottomrule
    \end{tabular}
    \caption{Median migration distance for stars of different ages and guiding radii. The migration distance is defined as $\rm R_{guide} - R_{birth}$, so a negative value means that stars of a given age (row) and guiding radius (column) had overall inward migration. We find younger stars have experienced less migration, and that stars located in the inner disk primarily migrated inwards. For our entire sample, we find that a median migration distance of 0.8 $\pm$ 2.6 kpc, and a radius change root mean square of 2.7 kpc.}
    \label{tab:migration}
\end{table*}

With estimates of stellar birth radii in the MW disk, we can probe the migration strength of the MW over time. Table \ref{tab:migration} presents our recovered median migration distance as a function of stellar age and guiding radius, where a negative value corresponds to inward migration (i.e. $\rm R_{birth} > R_{guide}$). We find a net outward median migration distance of 0.8 kpc for our sample, with more migration for older stars (in agreement with the effect of heating found in simulations; \citealt{Minchev2014}). The radius change root mean square of our sample ($\rm \sqrt{mean(R_{guide} - R_{birth})^2}$) is 2.7 kpc. While this value is in agreement with migration strengths reported in simulations \citep{2008roskarb, 2021Silva_simulation, Khoperskov2021} and other MW disk studies \citep{Frankel2020}, we stress that there is degeneracy in this value, as the causes of migration and their relative proportions can be different.

Similar to the 3 epochs found in the present-day gradient measured using guiding radii (Section \ref{sec:results_mw}), we also see 3 phases in the median migration strength with our estimated stellar birth radii. For stars with ages $\lesssim 6$, we find minimal ($<1$ kpc) net outward migration and smaller root mean square change in radius ($< 2.6$ kpc). This is a direct result of the more quiescent picture we described above, where the difference between the present-day gradient and recovered birth gradient only minimally vary. Thus, the migration found here is probably driven by internal perturbations. From $6-10$ Gyr ago, we find slightly more migration than in the more recent times (root mean square change in radius of $3.2 - 3.4$ kpc). This extra migration period burst may be due to Sagittarius, or, since it has been shown that most migration happens within a few Gyr of beginning of bar formation \citep{Minchev2011, Khoperskov2020}, our results could suggest that the MW bar formed $\sim 8-10$ Gyr ago (in agreement with other independent findings; e.g. \citealt{Bovy2019, Haywood2024, Sanders2024}). The oldest stars of our sample experience the most migration (root mean square radius change 4.3 kpc). Disentangling the sources of this migration is impossible, as it can be due to many different causes over time (GSE, Sagittarius, bar, spiral arms).

Across $\rm R_{guide}$, we find net inward migration for stars currently located in the inner disk, and net outward migration for stars $\rm R_{guide} > 5$ kpc, as expected from numerical simulations (e.g., \citealt{Minchev2014}, Fig.3). We estimate that $\sim 30\%$ of the stars in $\rm 3 < R_{guide} < 5$ kpc are inward migrators --- which we define as $\rm R_{birth} - R_{guide} > 2$ --- whereas the outer disk ($\rm 9 < R_{guide} < 11$ kpc) contains $\sim 30\%$ outward migrators. Looking across all ages, stars currently located in the solar neighborhood ($\rm 7 < R_{guide} < 9$ kpc) have, on average, experienced the most net migration. However, this conclusion is a bit misleading, as most mono-age bins show more net migration for larger $\rm R_{guide}$ (in agreement with simulations; \citealt{Renaud2024}). Therefore, we suggest that readers focus on the migration distance as a function of both age and guiding radius, and not marginalize over age.

\begin{figure*}
    \centering
    \includegraphics[width=0.5\textwidth]{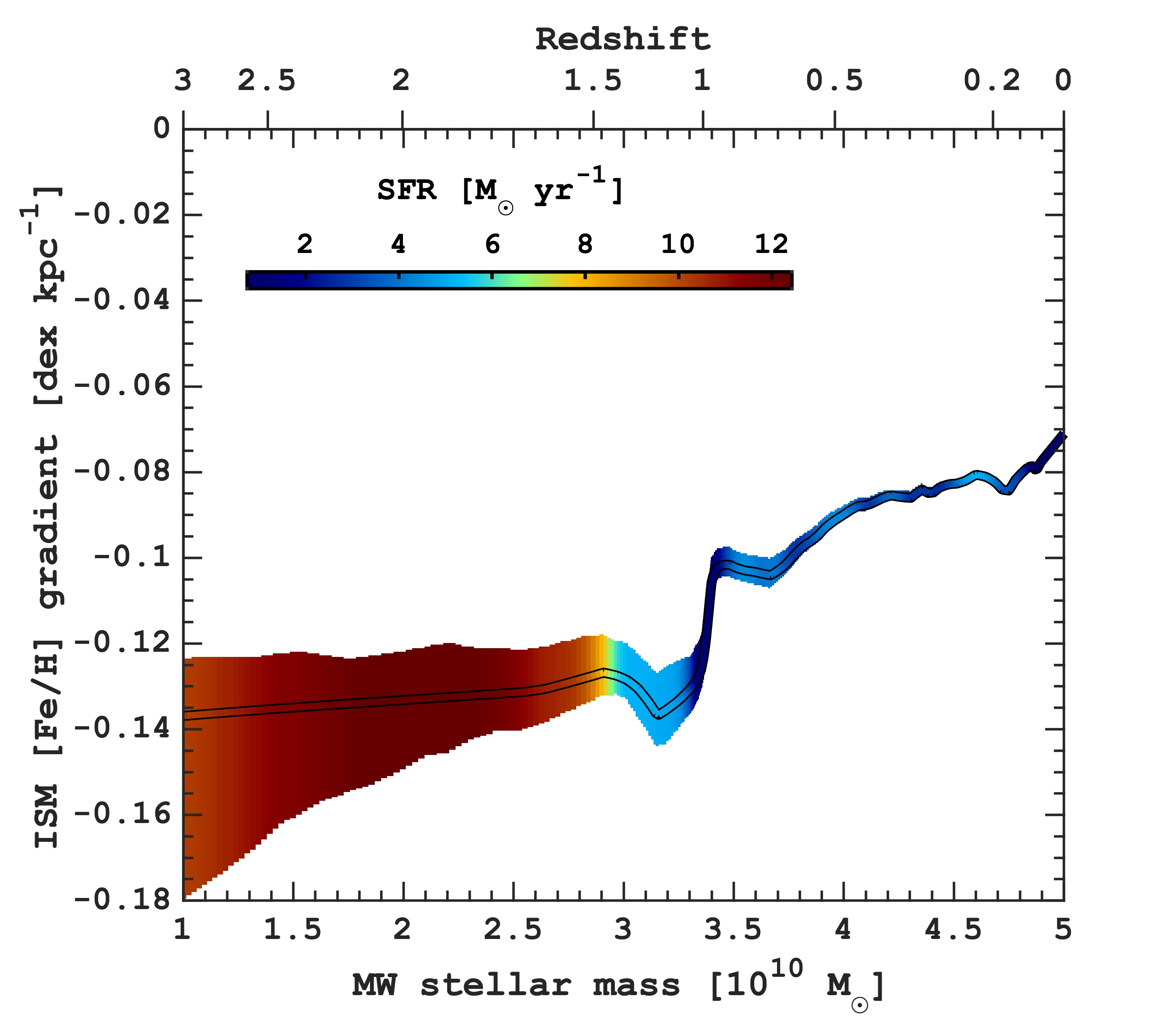}\includegraphics[width=0.5\textwidth]{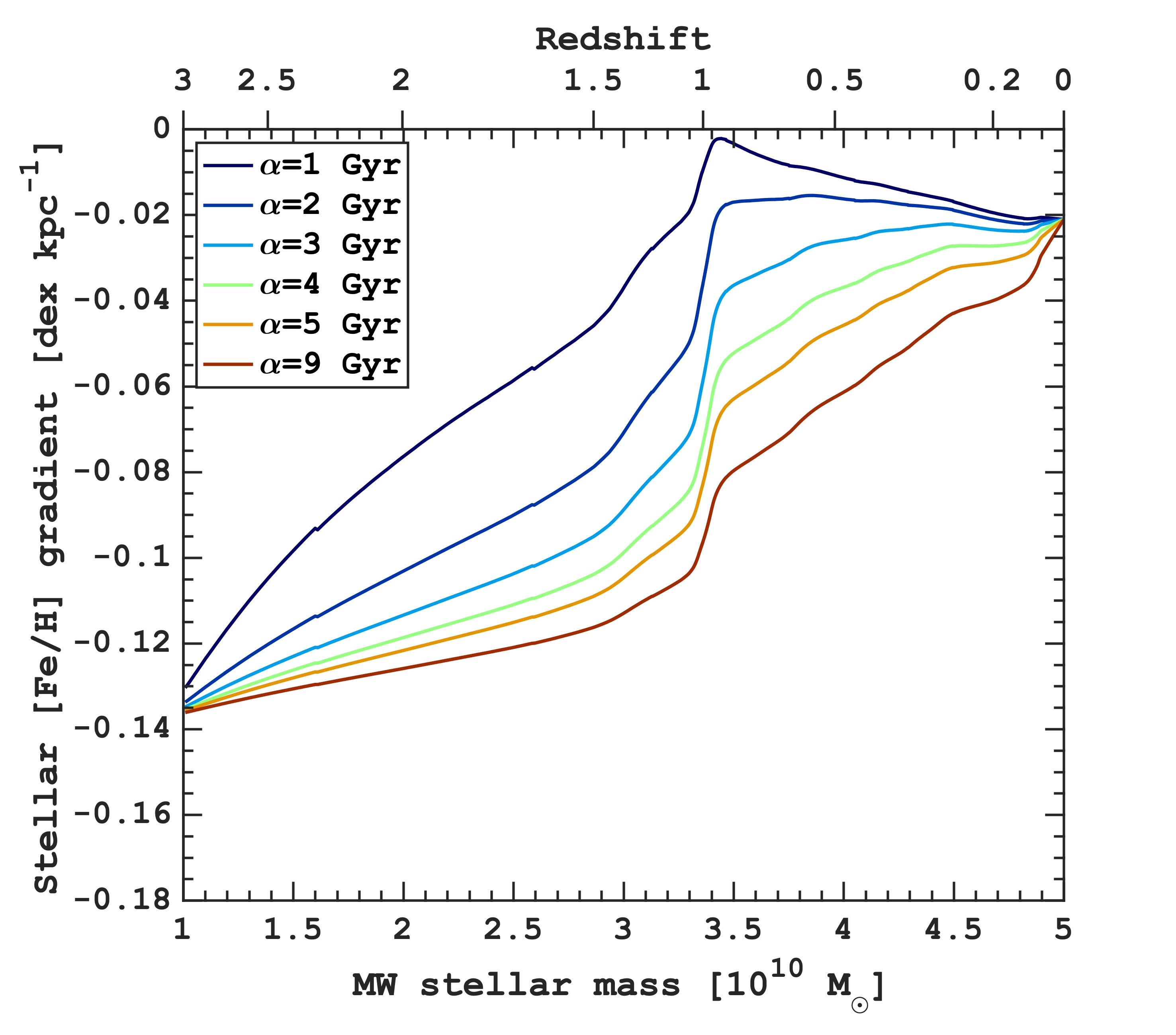}
    \caption{Evolution of the mass-metallicity gradient relation in the MW. {\bf Left:} evolution of the ISM radial metallicity gradient as a function of the MW stellar mass. The vertical width of the filled area is limited by the birth metallicity gradient uncertainty~(see Fig.~\ref{fig:mw_grad}), where the mean value is shown within the black lines. The color of the filled area corresponds to the star formation rate, which, together with the stellar mass growth of the MW, was adapted from \cite{2015A&A...578A..87S}.  {\bf Right:} evolution of the stellar mass-weighted radial metallicity gradient for different gradient evolution timescales, $\alpha$~(see Section~\ref{sec::MW_stellar_gradient_evolution} for more details). The analysis of the TNG50 galaxies~(see Fig.~\ref{fig::gradient_evolution_model}) and the relatively short active phase of stellar radial migration action support the rapid transformation of the mono-age stellar gradient, favouring the MW stellar metallicity gradient evolution with $\alpha=1-2$~Gyr. }\label{fig::gradients_evolution}
\end{figure*}

\section{Putting the evolution of metallicity gradients in the MW disk into context}\label{sec::gradients_in_time}
\subsection{ISM metallicity gradient}\label{sec::gradients_in_time_ISM}

In this section, we use the derived birth radii of MW disk stars to constrain the evolution of radial metallicity gradients up to redshift $\sim 3$. The solid black and blue lines in Fig.~\ref{fig:mw_grad} illustrate the birth radial metallicity gradient as a function of stellar age, which is equivalent to the evolution of the radial metallicity gradient in the star-forming ISM. Similar quantities are measured in large amounts for extragalactic systems, where the negative gradients dominate in disc galaxies at low redshifts~\citep{Ho2017a,Ho2017b}. However, the data for $z>1$ show a diverse picture with a large scatter from negative to flat and even positive gas metallicity gradients~\citep{2014MNRAS.443.2695S, 2016ApJ...827...74W, 2016ApJ...820...84L,Venturi2024}. Simulations somewhat allow to resolve this complexity, suggesting that strong negative metallicity gradients mostly appear in galaxies with a gas disc, as reflected by well-ordered rotation~\citep{Ma2017, 2012MNRAS.425..969P, 2013A&A...554A..47G}. Since the MW disc is believed to assemble quite early~($z > 3$; ~\citealt{2022MNRAS.514..689B, XiangRix2022, Semenov2024}), we may expect that it has a steep metallicity gradient. We note that this assumption is behind the birth radii calculation~\citep{Minchev2018_rbirth}, trivially requiring the existence of the disc component. The birth radii determination assumes that stars in our sample correspond to the disk populations; however, contamination of accreted stars, especially for stars older than $8-10$~Gyr is known~\citep{2019A&A...632A...4D,2020MNRAS.494.3880B}. The disentangling of accreted from in-situ~(genuine disk) stars is quite complex and not a fully resolved problem~\citep{2022MNRAS.510.2407B,2023arXiv231005287K,2024MNRAS.527.2165C}; however, we can safely neglect a small possible fraction of ex-situ populations as we do not use stars with metallicities $<-1$ and eccentricity $>0.5$ for the birth radii and abundance gradients calculation.

To make meaningful comparisons with extragalactic observations, it is essential to ensure that the gradients are compared to MW analogues at corresponding redshifts. In the left panel of Figure~\ref{fig::gradients_evolution}, we show the evolution of the radial metallicity gradient in the star-forming ISM as a function of MW stellar mass, which was adapted from \cite{2015A&A...578A..87S}, who recovered the MW mass-growth using chemical abundance patterns in the disk. Since the correlation between metallicity and star formation activity is expected~\citep{2010MNRAS.408.2115M,2013ApJ...772..119L}, the mass-metallicity gradient relation is color-coded with the star formation rate value at the corresponding epoch, providing an additional dimension for identifying the MW analogues. 

The evolution of the star-forming gas metallicity gradient in the MW is quite monotonic~(see left panel of Figure~\ref{fig::gradients_evolution}), in general agreement with a flattening over time predicted in chemical evolution models \citep{1997ApJ...475..519M,1998A&A...334..505P, 2000A&A...362..921H}. There is little gradient variation from $z \sim 3$ to $z \sim 1$ where it remains rather steep~($\rm \approx -0.14~ dex/kpc$). This epoch is characterised by the most intense phase of star formation when about half of the MW stellar mass was formed. 
Our birth radii suggest that the star-forming region is likely confined to the inner $\approx 6$~kpc during this time (Figure \ref{fig:SNdist}). The steep metallicity gradient during this evolutionary phase is somewhat surprising, as the high star formation rate would imply a high rate of turbulence and mixing, effectively erasing the radial structure of the metallicity distribution~\citep{2004ApJ...612..894B,2005ApJ...630..298B,2009ApJ...707L...1B}. \cite{Ma2017}, however, suggest that the metallicity gradient remains flat in such a bursty star formation regime if the galaxy remains clumpy with little rotation, which is not the case in the MW~\citep{2022MNRAS.514..689B, Khoperskov2023, 2022arXiv220402989C, 2024ApJ...962...84S}. Using the VINTERGATAN simulation~\citep{Agertz2021_vintergatanI, Renaud2021_vintergatanII, Renaud2021_vintergatanIII}, \cite{2022MNRAS.516.2272S} showed that a high rate of star formation is possible once the rotationally-supported disc is settled. Therefore, the current state of the debate about the existence (and the strength) of the radial metallicity gradient at early phases of the MW-like galaxies remains open, as different galaxy formation models and subgrid physics can provide us with different solutions. At the same time, the extragalactic observations favour a steep gradient at high redshift for already rotationally-supported galaxies~\citep{2013ApJ...765...48J,2015AJ....149..107J}.

The next feature in the gas metallicity gradient evolution is the non-monotonic change around $z\sim 1.5-1$~(see Fig~\ref{fig::gradients_evolution}, left panel). As already anticipated, this epoch corresponds to the infall of the GSE~\citep{Belokurov2018,Haywood2019} and simultaneously the end of the high-$\alpha$, inner disc formation~\citep[see, e.g.][]{2013A&A...560A.109H, Haywood2019, 2014A&A...562A..71B}. The small steepening of the gradient is rather short-lived with a rapid transition to a flattened gradient, in agreement with the merger-induced gradient transformation~\citep{Buck2023, Annem2024, Renaud2024}.

The most recent gradient evolution, during $z\sim 0.9-0$~(see left panel of Fig~\ref{fig::gradients_evolution}), is characterized by a slow flattening of the ISM metallicity gradient with the endpoint of $\rm \approx -0.07~dex/kpc$. Assuming the present-day half-mass radius~($\rm r_{eff}$) of the MW of $\approx 5$~kpc~\citep{2016ARA&A..54..529B}, we find the present-day ISM metallicity gradient scaled by the effective radius of $\rm \approx~-0.35~dex/r_{eff}$ in agreement with observational data from \cite{2013A&A...554A..58S,2015A&A...581A.103G}. As the star formation activity is rather low ~($\rm SFR\sim 1-2 M_\odot yr^{-1}$), not much ISM mixing is expected, and the flattening due to non-local enrichment by migrating stars is rather low. Although it is hard to estimate the contribution of the latter, we need to mention another source of the gradient flattening associated with the growth of the MW disc in size. TNG MW/M31-like galaxies show a push in the star-forming region outwards~(see Fig. 6 in \citealt{Ratcliffe2024_tng50}), effectively increasing the size of the galaxies. At the same time, the extragalactic observations suggest the rapid growth of galaxies in size from $z\sim 1$ ~\citep{2024A&A...682A.110B}. Therefore, we can expect that the detected flattening of the ISM gradient is partially linked to the growth of the MW disc in size, implying the build-up of its outer disc.

\subsection{Stellar metallicity gradient}\label{sec::MW_stellar_gradient_evolution}
Another potential application of the data obtained using the birth radii reconstruction is the possibility to explore how the total stellar metallicity gradient evolved in the MW. Since we know the birth~(or ISM) and the present-day gradients for mono-age populations, we need to assume how the masses of these populations and gradients evolve with time between these two points. Mass evolution can be obtained pretty straightforwardly. For each mono-age population, we adopt the initial mass from the MW star formation history~\citep{2015A&A...578A..87S}. Then, the mass evolution of each mono-age population is governed by the mass loss by massive stars, which we calculate using chempy~\citep{2017A&A...605A..59R}\footnote{https://github.com/jan-rybizki/Chempy} assuming a single burst model, Chabrier initial mass function~\citep{2003PASP..115..763C} and the contribution from SNI~\citep{2013MNRAS.429.1156S}, SNII~\citep{2013ARA&A..51..457N} and AGB stars~\citep{2010MNRAS.403.1413K}. As a result, we obtained the mass evolution for mono-age populations accounting for the stellar mass loss~(winds) of stars and stars dying in the SNe. We note that the high uncertainties in stellar ages, MW star formation history, and other parameters make no big difference if we choose different sets of yields and initial mass functions. Under any reasonable assumption, the stellar mass loss happens quite rapidly, yielding about $40-50\%$ of the initial mass after $1-2$ Gyr since the mono-age population formation with a negligible amount of the mass loss at later times. 

The evolution of the metallicity gradient for a mono-age population with the known starting and end points is not trivial. In Figure~\ref{fig:mw_grad}, we showed that the initial~(at birth) gradient is always steeper than the final one~(present-day) for all ages, suggesting a monotonic flattening. This aligns with the assumption about the evolution of the mono-age stellar population gradient governed by radial migration. Churning~(change of the angular momentum) disk stars by spiral arms is most efficient for young populations while they are on nearly-circular orbits~\citep{Selwood2002,Roskar2008_migration}; once stars are heated, the probability for them to be churned decreases. Hence, we can assume that the metallicity gradient of mono-age populations also evolves on a short time scale. In our toy model, we assume that the evolution of the mono-age stellar radial metallicity gradient follows the following expression:
\begin{multline*}
\rm     \nabla [Fe/H] (t) = \nabla [Fe/H] (t_{birth}) + (\nabla [Fe/H] (t_{now}) - \\ \rm \nabla [Fe/H](t_{birth}))\times \Theta(t)\,, 
\end{multline*}
where
$$
\rm     \Theta(t) = \frac{erf((t - t_{birth)})/(\sqrt{2} \alpha))}{erf((0 - t_{birth)})/(\sqrt{2} \alpha))}\,. 
$$
These expressions parameterize the flattening of the gradient with time and are scaled by the initial and the present-day values. The shape of the function is controlled by the $\alpha$ parameter, which indicates how rapidly the gradient transforms. The shape of the corresponding functions for $\alpha=1$~Gyr is presented in Figure~\ref{fig::gradient_evolution_model}~(right) where the colored lines reflect different possible evolutions of stellar metallicity gradients for different mono-age populations with the starting~(birth) and final~(present-day) points taken from Figure~\ref{fig:mw_grad}. The choice of particular parameterization does not play a significant role, as it only needs to provide a monotonic transition from initial to final gradient value with a single parameter controlling the time scale of transition.

In order to verify how reasonable the adopted model (right panel of Figure~\ref{fig::gradient_evolution_model}) is, we present the evolution of the mono-age stellar gradient evolution in four TNG50 galaxies in our sample. As one can see, the parameterization we use agrees nicely with the mono-age metallicity gradients evolution in terms of the global shape, highlighting a rapid change on a scale of $1-2$~Gyr with negligible evolution later on. The TNG50 galaxies show a number of short time-scale variations and non-monotonic behaviour (see also \citealt{Renaud2024} for mono-age gradient evolution in VINTERGATAN); however, the global shape of the curves assures us with the choice of the model we made. 

Once we have the mass of the MW mono-age stellar populations together with their individual evolution of the metallicity gradient, we can calculate the mass-weighed metallicity gradient evolution. In the right panel of Figure~\ref{fig::gradients_evolution} we present the total stellar metallicity gradient evolution as a function of the MW stellar mass and redshift. The lines of different colours start and end at the same points, corresponding to the metallicity gradient of the oldest mono-age populations and the present-day mass-weighted stellar metallicity gradient of $\rm \approx -0.02~dex/kpc$~($\rm \approx -0.1~dex~r^{-1}_{eff}$), which is similar to the gradients found in the nearby galaxies of similar type~\citep{2014A&A...570A...6S,2017MNRAS.465.4572Z} The colour of the lines corresponds to the choice of the $\alpha$ parameter~($1-5$ and $9$ Gyr), governing how rapidly the mono-age gradient changes from the initial to the present day. The range of $\alpha$ values is chosen to demonstrate the full range of the possible evolution of the stellar metallicity gradient. Although, the $\alpha=9$~Gyr seems to be quite unrealistic, suggesting that the oldest stars have transformed their gradient recently, $1-2$~Gyr ago. In this case, the steep gradient could be observed until $ z\sim 1$ with the following rapid flattening, which is hard to understand if it were driven by the radial migration associated with cold stellar populations. 

On the other hand, in models with the rapid mono-age gradient evolution~($\alpha=1$ or $2$~Gyr), we can see quite an interesting picture. The stellar metallicity gradient flattens rapidly~($z\sim 3-1$, over $\approx 3.5$~Gyr) to the present-day value or even a zero gradient~($\alpha=1$). As this epoch corresponds to the end of the high-$\alpha$ sequence formation, we suggest that the stellar metallicity gradient was negligible by the end of the inner disc formation, which is somewhat different from the fully mixed ISM assumption during this epoch~\citep{2015A&A...579A...5H, 2018A&A...618A..78H}, which as we discussed in the previous section may not be applicable if the MW was already rotationally-supported (and thus implying a negative metallicity gradient). In any case, our model suggests the lack of the stellar metallicity gradient at $z \approx 1$. The following evolution of the gradient is driven by the quiescent low-$\alpha$ disc formation phase; those mono-age populations, even forming on a steeper metallicity gradient, weakly affect the total stellar metallicity gradient. 

For completeness, in Fig.~\ref{fig::gradients_evolution}~(right panel) we show the evolution of the total stellar gradient for $\alpha=3-5$~Gyr, revealing an intermediate evolution between $\alpha = 1$ and $\alpha = 9$~Gyr. However, as we argued above, we favour the models with lower $\alpha$, which are more in line with our argumentation about the radial migration process and the gradients evolution in simulated galaxies~(see Fig.~\ref{fig::gradient_evolution_model}). 

To summarize, we suggest that the present-day stellar metallicity gradient was established at $z \approx 1$, around the end of the inner disc, high-$\alpha$ formation, and it has not evolved significantly since then. The latter is the result of rather weak radial migration over this period of time~(see Sect.~\ref{sec::migration_strength}), and also by a modest increase in the stellar mass spread out over a larger area of the disc. 



\section{Conclusions}
\label{sec:conclusions}

The use of stellar birth radii is rapidly growing, allowing for a detailed view of the MW disk evolution with time that minimizes the effect of radial migration. Estimating $\rm R_{birth}$ requires knowledge of how the metallicity gradient evolves over time, which can be recovered through the scatter in [Fe/H] across age \citep{Lu2022_Rb}. However, as shown in \cite{Ratcliffe2024_tng50}, weakly barred galaxies exhibit a continual growth in the width of the star-forming region causing this method to fail, whereas strongly barred galaxies exhibit quenching in the bar region that grows with time allowing for the method to have better results. In this work, we proposed a correction to adjust for the time evolution of the width of the star-forming region, which allows for a more universal recovery of $\rm \nabla [Fe/H](\tau)$ and improves the ability to recover the metallicity gradient to within 20\%. Applying our correction to the MW provides insight into its evolution, which we put into context to allow for comparisons to external galaxies. Our main conclusions are as follows:

\begin{itemize}
    \item Our proposed correction --- which captures the growth of the star-forming region --- is dependent on bar strength (Figure \ref{fig:dr}), and appears to be independent of simulation prescriptions for stronger-barred galaxies (left panel of Figure \ref{fig:buck}). This growth creates a non-negligible effect on the correlation between $\rm \nabla [Fe/H](\tau)$ and the scatter in [Fe/H] for different age bins. For weakly barred galaxies, $\rm \Delta R(\tau$) can vary from galaxy to galaxy (Figures \ref{fig:dr} and the right panel of Figure \ref{fig:buck}), but there is an overall improvement on the relationship between the metallicity range and gradient after applying our proposed correction.
    

    

     \item In applying our proposed correction to the MW (Figures \ref{fig:mw_grad} and \ref{fig::gradients_evolution}), we find that the ISM metallicity gradient was as steep as $\rm -0.14~dex/kpc$ until $z\approx 1$, with a prominent non-monotonic variation at the time of the GSE infall and end of the inner high-$\alpha$ disc formation~($\approx 8$~Gyr ago). This is followed by a gradual flattening to $\rm \approx -0.07~dex/kpc$~($\rm \approx -0.35~dex~r^{-1}_{eff}$) at present. 

    \item The birth radii distributions of young stars at different Galactic radii always peak close to the current radius location, assuring the reliability of our method (Figure \ref{fig:SNdist}). Using this improved method, we find that the Sun formed at $6.66 \pm 0.58$ kpc. 

    \item We find that the migration rate varies with age and, as expected, it is higher for older stars. However, we observe a noticeable decline in migration rate for stars younger than $6-8$~Gyr showing little migration~($\lessapprox 1$~kpc) while the older stars migrate for more than $2$~kpc from their birthplaces. Such a sharp transition can not be explained by the accumulated migration, as its effect saturates with time. Therefore, we suggest that different migration mechanisms dominated in different epochs of the MW evolution~(such as interactions, spiral arms, bar formation).

    \item Using assumptions regarding the mono-age stellar metallicity gradient evolution, in agreement with simulated galaxies, we introduced a model for reconstructing the total stellar metallicity gradient evolution in the MW up to redshift $\approx 3$ (Figure \ref{fig::gradients_evolution}). Relying on the arguments about the radial migration efficiency, we favour models with a rapid mono-age gradient transformation which suggest a nearly flat stellar metallicity gradient since $z\approx 1$, of $<-0.03$ dex/kpc~($\rm \approx -0.1$ dex $\rm r^{-1}_{eff}$).


\end{itemize}

Our results illustrate that accounting for the growth of the star-forming region in a galactic disk is mandatory to accurately recover $\rm \nabla [Fe/H](\tau)$ from Range[Fe/H](age). Our recovered mono-age birth radii distributions for APOGEE DR17 red giant disk stars with ages from \texttt{distmass} are reassuring that our correction indeed works on the MW disk, and provides better $\rm R_{birth}$ estimates. To choose the optimal steepest gradient strength, we minimize $\rm |R - R_{birth}|$ for stars with ages $<2$ Gyr as we expect these stars have not had the opportunity to move far away from their birth sites. However, nearly 2 Gyr old stars may have had enough time to net migrate outwards (e.g. \citealt{Lian2022}). Therefore we acknowledge that in an ideal setting we would be minimizing the migration strength for only the extremely young stars ($< 0.5$ Gyr or so), but due to age catalog limitations we have to minimize over this $0-2$ Gyr age group. Minimizing over different criteria changed the reported steepest gradient and subsequent migration strengths, but the shape of the recovered gradient (Figure \ref{fig:mw_grad}) and trends with migration strength (Table \ref{tab:migration}) remained in agreement. If more net outward migration is required for the $0-2$ Gyr age group, then the steepest gradient would decrease to $\sim -0.15$ dex/kpc, depending on how much migration is warranted.


The proposed correction to the range in [Fe/H] across age (Section \ref{section:method_correction}) was applied in this work without stacking the galaxies \citep[e.g.][]{Cheng2024}. While we find stacking does not alter our results, we choose to provide the non-stacked correction as the growth of the galactic disks are naturally already accounted for in the bar strength separation \citep[see][]{Khoperskov2023}.

The rapidly expanding body of research using stellar birth radii~\citep[e.g.][]{Minchev2018_rbirth, Frankel2018, Frankel2020, Lu2022_Rb, Ratcliffe2023_enrichment, Wang2023, Ratcliffe2023_chemicalclocks, Spitoni2024, Marques2024, Plotnikova2024} offers a valuable new opportunity to explore not only radial migration within the Milky Way (MW) disk, but also to impose tighter constraints on its historical evolution. Moreover, this research enables detailed comparisons with external galaxies up to redshift of $\approx 3$. With the forthcoming data from large-scale Galactic spectroscopic surveys such as 4MOST~\citep{deJong2019_4most}, WEAVE~\citep{Dalton2012_weave}, and SDSS-V~\citep{sdssV}, these constraints are expected to become even more precise. This will significantly enhance our understanding of the MW evolution, as well as that of its analogues, both in the nearby universe and at higher redshifts, filling the gaps in galaxy evolution theory in general.

\section*{Acknowledgements}

B.R. and I.M. acknowledge support by the Deutsche Forschungsgemeinschaft under the grant MI 2009/2-1. 

\nolinenumbers

\bibliography{Ratcliffe}{}
\bibliographystyle{aa}
\newpage

\appendix

\section{NPMLE}\label{sec:methods_npmle}

Each observed value $\rm (age_i, [Fe/H]_i) =$ $Y_i$ can be assumed to follow $$Y_i = \Theta_i + \epsilon_i \qquad \epsilon_i \sim N_d(0, \Sigma_i)$$ where $\Sigma_i$ is the known measurement uncertainty $\forall\, i$ and $\Theta_i$ represents the true (age, [Fe/H]) value for star $i$. If we assume the $\Theta_i$'s are i.i.d. from a prior distribution $G$ and are independent with the $\epsilon_i$'s, then the $Y_i$'s can be modeled as $$ Y_i|\Theta_i = \theta \sim N_d(\theta,\Sigma_i).$$ Our main goal is to denoise our observed values $Y_i$, and try to recover the true $\Theta_i$. Depending on the structure of the unknown $\Theta_i$'s, it is possible to get significant improvements over the typically used estimator $Y_i$ by using information from the other stellar measurements ($Y_j,\, j \neq i$) in addition to $Y_i$ to estimate $\Theta_i$. An example of this could be if you have a star with an age estimate of 15 Gyr, the age estimates of the other stars (which would be $<13$ Gyr) could help provide a more reliable estimate. In order to use this additional information, we use the oracle Bayes estimator: $$\theta_i^* = E(\Theta_i^*|Y_i) \qquad \Theta^* \sim G \qquad Y_i|\Theta^* = \theta \sim N_d(\theta, \Sigma_i),$$ which can be rewritten as $$\theta_i^* = E(\Theta^*|Y_i) = Y_i + \Sigma_i \frac{\nabla f_{G, \Sigma_i}(Y_i)}{f_{G, \Sigma_i}(Y_i)},$$ where $f_{G, \Sigma_i}(Y_i)$ is the marginal density of $Y_i$. In order to feasibly get an estimate of $\theta_i^*$, we need to estimate $G$. The NPMLE of $G$ is given by $$\hat{G} \in \underset{G}{\text{argmax } }\Pi_{i=1}^n f_{G, \Sigma_i}(Y_i).$$ After solving for $\hat{G}$, the denoised values can be solved as the posterior means. We refer the reader to \cite{Soloff2021_npmle} for more information.

\section{Extra Figures}

\begin{figure*}
     \centering
     \includegraphics[width=.9\textwidth]{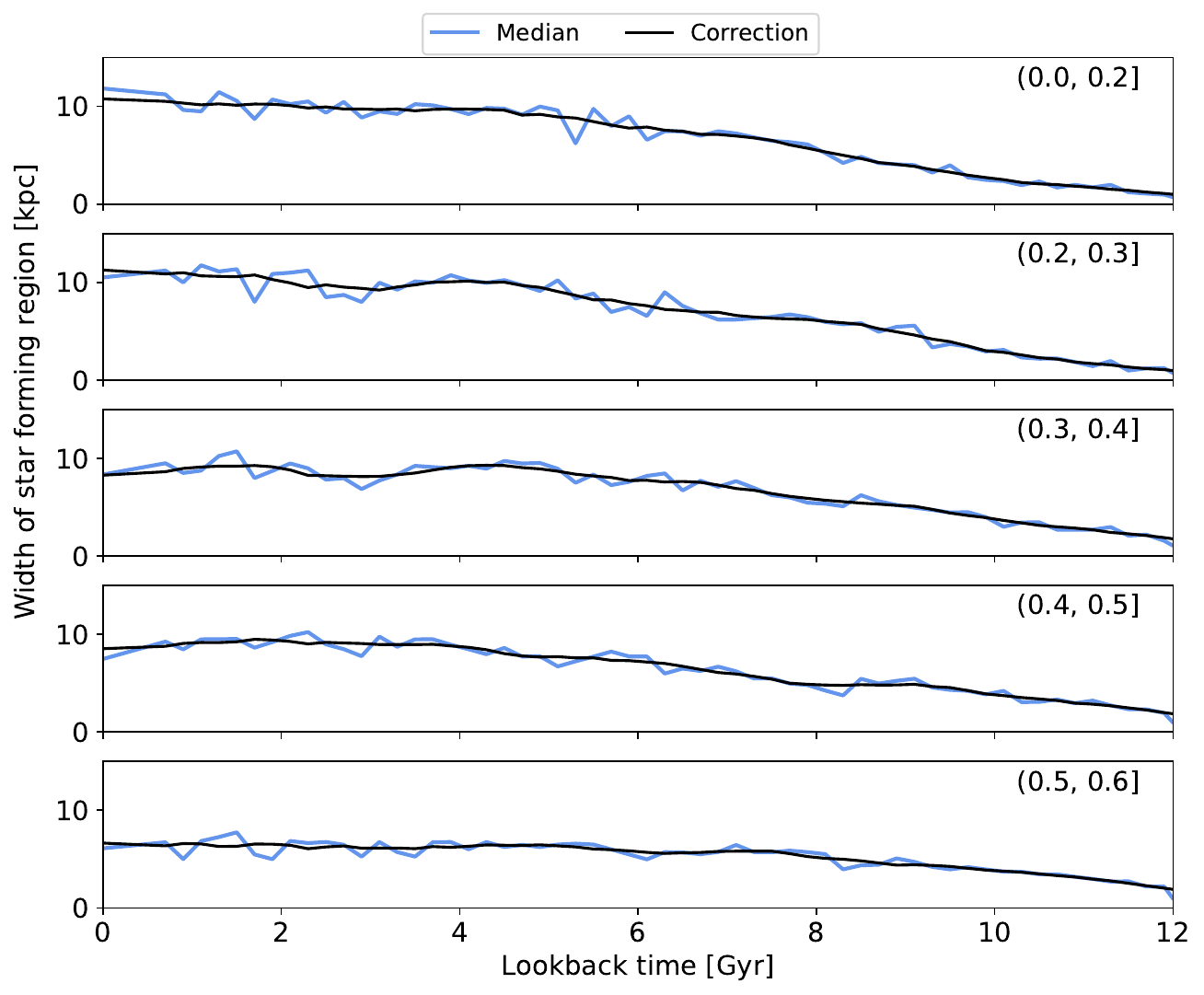}\\
\caption{The proposed correction to adjust for the width of the star-forming region (black line) for each present-day bar strength group. The blue line is the median width of the star-forming region, measured every 0.2 Gyr. The correction is taken as the running mean of the blue line over 1.2 Gyr.} 
\label{fig:correction}
\end{figure*}

Figure \ref{fig:correction} shows the median width of the star-forming region as a function of lookback time measured every 0.2 Gyr (blue lines). The correction we propose is taken as the running mean over 1.2 Gyr (black lines).

\begin{figure*}
     \centering
     \includegraphics[width=.32\textwidth]{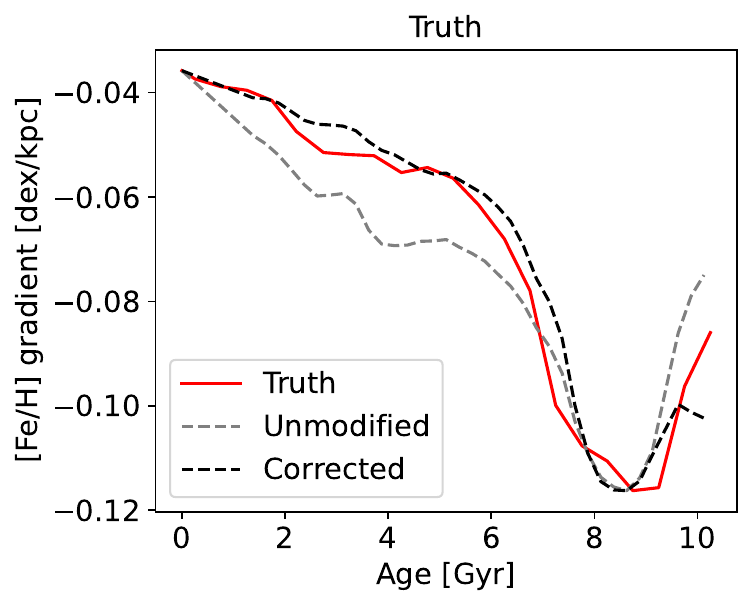}
     \includegraphics[width=.32\textwidth]{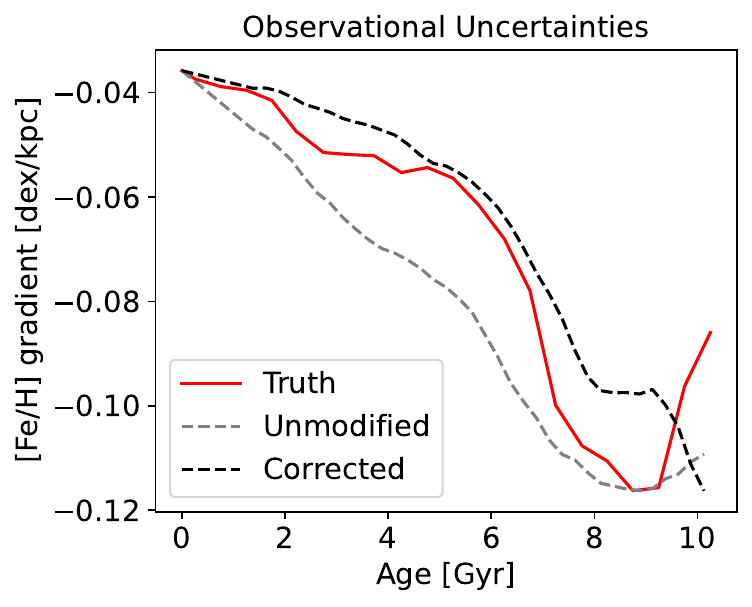}
     \includegraphics[width=.32\textwidth]{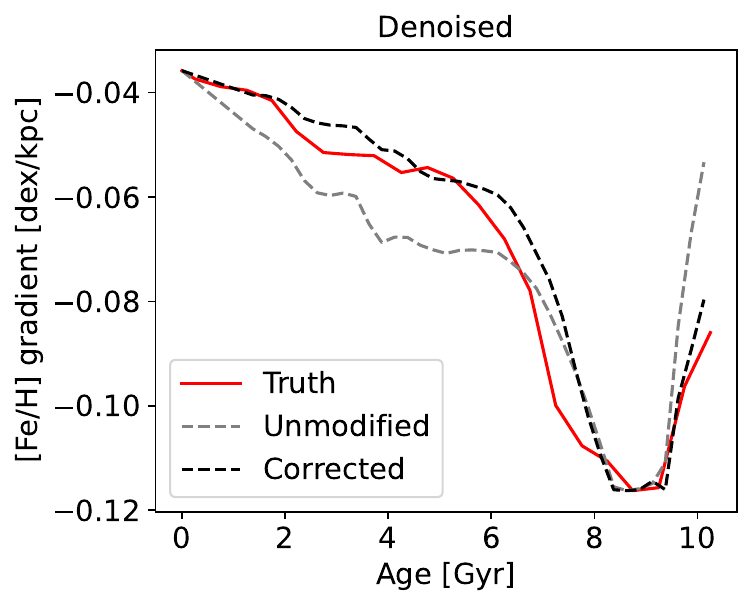}\\
\caption{{\bf Left:} Ability to recover $\rm \nabla [Fe/H](\tau)$ from Range[Fe/H](age) in g2.79e12 without any observational errors added. {\bf Middle:} Same as left but with 15\% age and 0.05 dex [Fe/H] uncertainties added. The range becomes artificially high at older ages past the knee in the age--metallicity relation due to present-day age errors. {\bf Right:} Recovery of $\rm \nabla [Fe/H](\tau)$ using the denoised age-metallicity relation. The comparison between denoising the data and the truth (left panel) is strikingly similar.} 
\label{fig:buck_grads}
\end{figure*}

Figure \ref{fig:buck_grads} demonstrates the success using the correction provided in this paper. The left panel shows the true (red line) and recovered metallicity gradients (grey and black dashed lines) in NIHAO-UHD galaxy g2.79e12 with no uncertainties added to (age, [Fe/H]). The recovered $\rm \nabla [Fe/H](\tau)$ using the method proposed in \cite{Lu2022_Rb} does a reasonable job of capturing the shape of the gradient, however it overestimates the gradient at younger ages and underestimates at older ages. Correcting for the time variation of the width of the star-forming region (black dashed line) provides a much better recovery. The middle panel illustrates the necessity of denoising the age--metallicity relation in the presence of present-day observational uncertainties. Once uncertainties are added, the recovered $\rm \nabla [Fe/H](\tau)$ is too steep at older ages, and the gradient is too smooth. The right panel shows that after denoising, the recovery is similar to when no uncertainties were added (such as in the left panel).

\begin{figure}
     \centering
     \includegraphics[width=.45\textwidth]{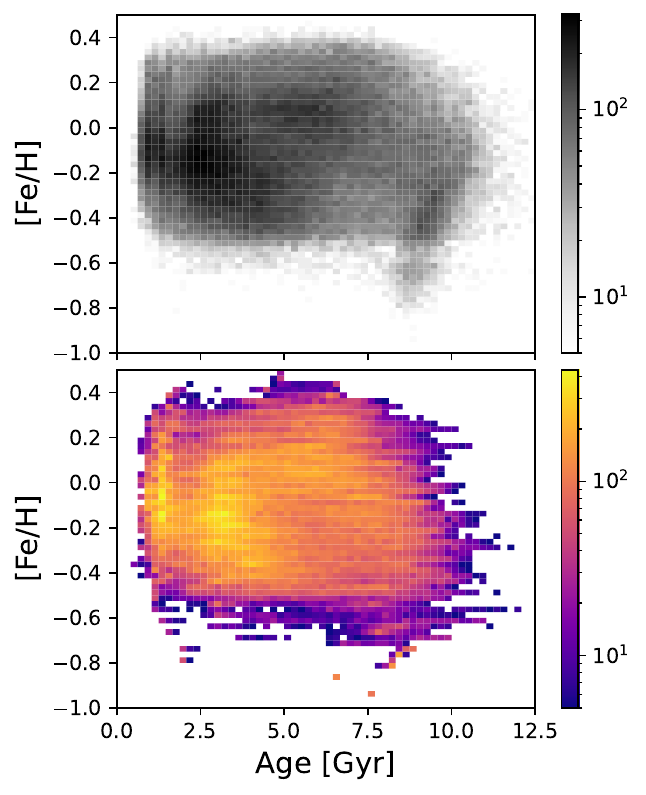}\\,
\caption{ {\bf Top:} Density distribution of the age--metallicity relation for our APOGEE disk sample with \texttt{distmass} ages. {\bf Bottom:} The recovered age--metallicity relation after denoising for one of the 100 Monte Carlo samples.} 
\label{fig:apogee_denoise}
\end{figure}

The top panel of Figure \ref{fig:apogee_denoise} provides the age--metallicity relation for our APOGEE sample. To have the best understanding of the evolution of our Galaxy, we redraw new ages and metallicity measurements for each star before denoising to estimate the error on the recovered gradient. The bottom panel of Figure \ref{fig:apogee_denoise} illustrates an example of the denoised age--metallicity relation.

\begin{figure*}
     \centering
\includegraphics[height=.38\textwidth]{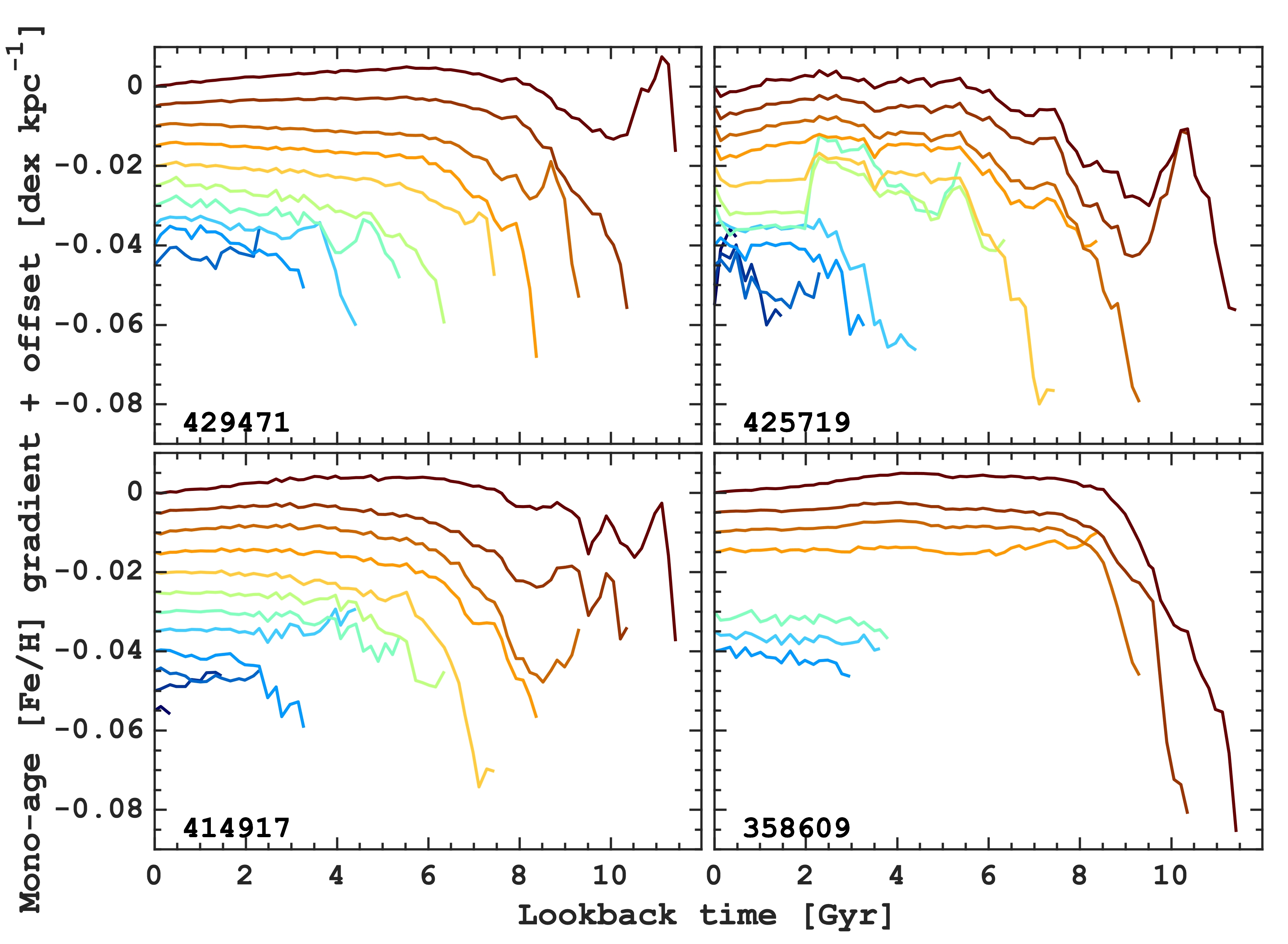}
\includegraphics[height=.39\textwidth]{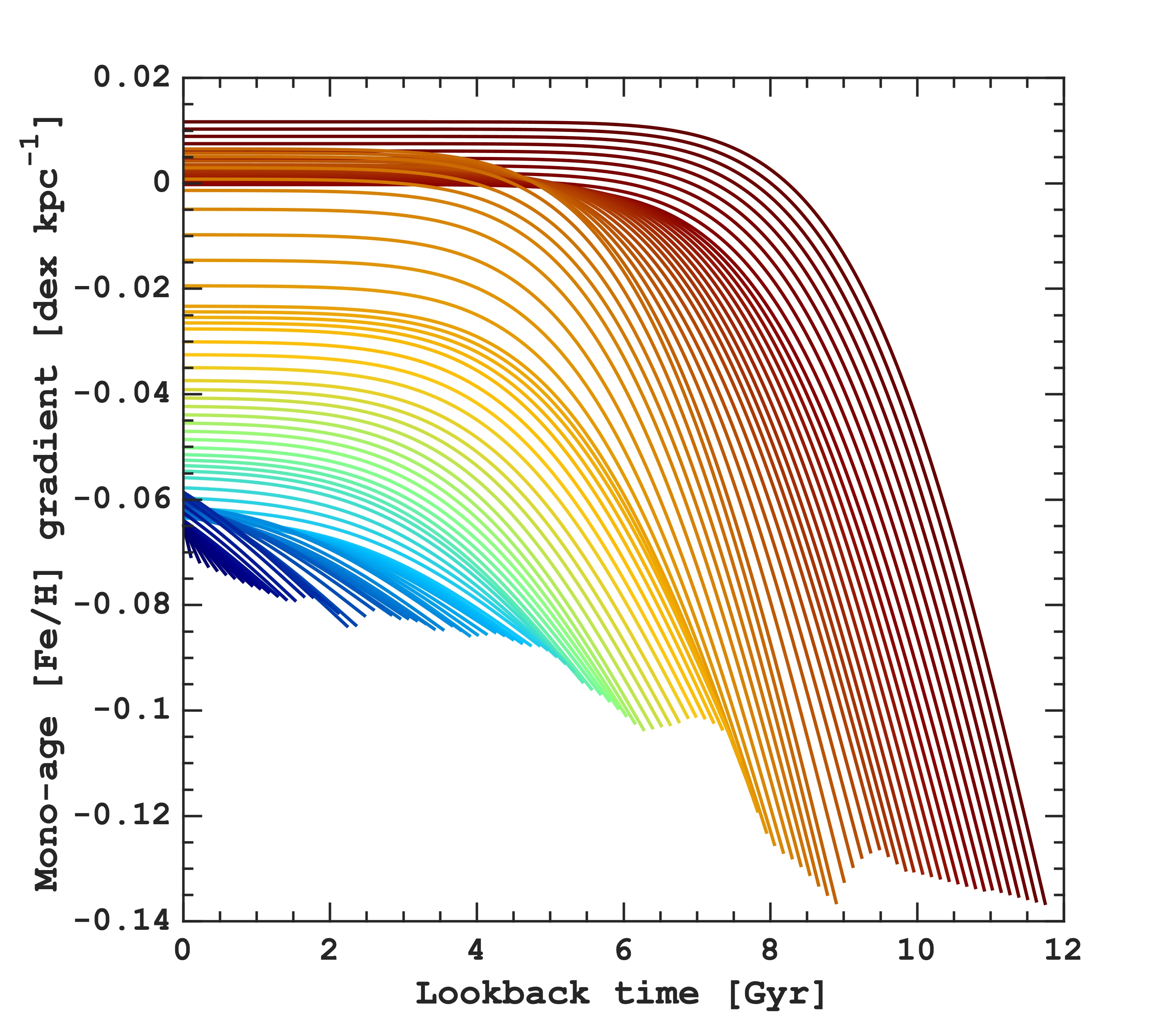}
\caption{{\bf Left:} evolution of the radial metallicity gradient for mono-age stellar populations in four barred TNG50 galaxies. {\bf Right:} assumed evolution of the mono-abundance metallicity gradients in the MW. In both panels, the lines of different colours correspond to populations of different ages, where red represents older aged populations and blue shows the evolution for younger aged populations. In the right panel, the initial and present-day gradient values are obtained in this work and presented in Fig.~\ref{fig:mw_grad}. The mono-age metallicity gradient evolution model is described in Section.~\ref{sec::MW_stellar_gradient_evolution} with the time-scale parameter $\alpha=1$~Gyr.} 
\label{fig::gradient_evolution_model}
\end{figure*}

The left panel of Figure \ref{fig::gradient_evolution_model} shows the evolution of the radial metallicity gradient in four TNG50 MW/Andromeda-like galaxies. The colored lines correspond to different mono-age populations. The right panel shows the results of how a mono-age gradient evolves over time, estimated from our Model in Section \ref{sec::gradients_in_time}.


\end{document}